\documentclass[floatfix,a4paper,aps,prd,nofootinbib,twocolumn,preprintnumbers]{revtex4-1}
\usepackage{amsmath}
\usepackage{amssymb}
\usepackage{amsfonts}
\usepackage{color}
\usepackage{axodraw4j}
\usepackage{graphicx}
\usepackage{subfigure}
\usepackage{geometry}
\usepackage[applemac]{inputenc}
\usepackage{geometry}
\usepackage{url}
\usepackage{hyperref}

\newcommand{\newc}{\newcommand}

\def\slashchar#1{\setbox0=\hbox{$#1$}     		
   \dimen0=\wd0                                 	
   \setbox1=\hbox{/} \dimen1=\wd1               	
   \ifdim\dimen0>\dimen1                        	
      \rlap{\hbox to \dimen0{\hfil/\hfil}}      	
      #1                                        	
   \else                                        	
      \rlap{\hbox to \dimen1{\hfil$#1$\hfil}}   	
      /                                         	
   \fi}

\newc{\eg}{{\it e.g.}\ }
\newc{\ie}{{\it i.e.}\ }
\newc{\cf}{{\it cf.}\ }
\newc{\idem}{{\it id.},\ }

\newc{\tev}{\mbox{~TeV}}
\newc{\gev}{\mbox{~GeV}}
\newc{\Mev}{\mbox{~MeV}}
\newc{\pb}{\mbox{~pb}}
\newc{\fb}{\mbox{~fb}}
\newc{\invpb}{\mbox{~pb}^{-1}}
\newc{\invfb}{\mbox{~fb}^{-1}}
\newc{\Psix}{{\mathrm{P}_{\!6}}}
\newc{\nPsix}{{\not{P}_{6}}}

\newc{\tanb}{\tan\beta}
\newc{\etmiss}{\slashchar{E}_T}

\newc{\lam}{{\bf \lambda}}
\newc{\lamp}{{\bf \lambda}^{\prime}}
\newc{\lampp}{{\bf \lambda}^{\prime\prime}}

\newc{\slepton}{\tilde{\ell}}
\newc{\squark}{\tilde{q}}
\newc{\ssup}{\tilde{u}}
\newc{\ssdown}{\tilde{d}}
\newc{\ssstrange}{\tilde{s}}
\newc{\sscharm}{\tilde{c}}
\newc{\sstop}{\tilde{t}}
\newc{\ssbottom}{\tilde{b}}
\newc{\sse}{\tilde{e}}
\newc{\ssmu}{\tilde{\mu}}
\newc{\sstau}{\tilde{\tau}}
\newc{\ssnu}{\tilde{\nu}}
\newc{\ssnue}{\tilde{\nu}_{e}}
\newc{\ssnumu}{{\tilde{\nu}_{\mu}}}
\newc{\ssnutau}{{\tilde{\nu}_{\tau}}}
\newc{\ssbnue}{\bar{\tilde{\nu}}_{e}}
\newc{\ssbnumu}{\bar{\tilde{\nu}}_{\mu}}
\newc{\ssbnutau}{\bar{\tilde{\nu}}_{\tau}}
\newc{\neut}{{\tilde{\chi}}^0}
\newc{\charge}{\tilde{\chi}}
\newc{\glu}{\tilde{g}}
\newc{\Higgs}{H^0}
\newc{\smass}{\tilde{m}}

\newc{\ttext}[1]{{\color{blue}  #1}}
\newc{\htext}[1]{{\color{red}  #1}}

\begin{document}

\title{Bounds on R-parity Violation from Resonant Slepton Production at the LHC}

\author{H.~K.~Dreiner}
\email[]{dreiner@th.physik.uni-bonn.de}
\affiliation{Bethe Center for Theoretical Physics and Physikalisches 
Institut, Universit\"at Bonn, Bonn, Germany}

\author{T.~Stefaniak}
\email[]{tim@th.physik.uni-bonn.de}
\affiliation{Bethe Center for Theoretical Physics and Physikalisches 
Institut, Universit\"at Bonn, Bonn, Germany}

\begin{abstract}
We consider the \texttt{ATLAS} and \texttt{CMS} searches for dijet
resonances, as well as the \texttt{ATLAS} search for like--sign dimuon pairs
at the LHC with 7 TeV center of mass energy. We interpret their
exclusions in terms of bounds on the supersymmetric R-parity violating
parameter space. For this we focus on resonant slepton production
followed by the corresponding decay.

\end{abstract}

\preprint{BONN-TH-2012-01}

\maketitle

\section{Introduction}
\label{Sect:Intro}
After initial problems~\cite{Gedicht}, the LHC has been running very
well since Nov. 2009. One of the main physics objectives is to search
for new physics beyond the Standard Model of particle physics (SM), in
particular also supersymmetry (SUSY)~\cite{Nilles:1983ge}.  The \texttt{CMS}
and \texttt{ATLAS} experiments have so far mainly concentrated on
R-parity conserving supersymmetry searches~\cite{Farrar:1978xj}, where
the lightest supersymmetric particle (LSP) is stable as well as
electrically and color neutral.  The corresponding searches thus
employ strict cuts on the missing transverse energy, $\etmiss$ (MET)~\cite{ATLAS_SUSYMETsearches,CMS_SUSYMETsearches}. To-date no disagreement with the SM has been found,
resulting in strict lower mass bounds on the new supersymmetric
particles in the simplest supersymmetric models; see also~\cite{Bechtle:2011dm}.

R-parity violation is theoretically equally well motivated~\cite{Dreiner:1997uz,Allanach:2003eb,Dreiner:2003hw,Dreiner:2005rd,Hirsch:2000ef} 
to the R-parity conserving case. It has the same particle content and
the same number of imposed symmetries. In particular it automatically
includes light neutrinos~\cite{Hall:1983id,Davidson:2000uc,Dreiner:2010ye},
without adding a new see-saw energy scale or right-handed neutrinos~\cite{Minkowski:1977sc,Mohapatra:1979ia}.  If R-parity is replaced by
 baryon-triality~\cite{Dreiner:2005rd,Dreiner:2006xw,Lee:2010vj,Dreiner:2011ft},
the superpotential must be extended by
\begin{equation}
W_{B_3} = \lam_{ijk} L_i L_j \bar E_k + \lamp_{ijk} L_i Q_j \bar D_k 
+ \kappa_i L_i H_u\,,
\label{Eq:WnPsix}
\end{equation}
where we have used the notation as in~\cite{Allanach:2003eb}.
These operators all violate lepton number. At a hadron collider the
terms $\lamp_{ijk} L_iQ_j\bar D_k$ can lead to resonant slepton and
sneutrino production \cite{Dimopoulos:1988fr}
\begin{align}
\bar d_j + d_k &\to \ssnu_{Li} , \label{Eq:production1}\\
\bar u_j+ d_k &\to \slepton_{Li}^-\,,\label{Eq:production2}
\end{align}
as well as the charge conjugate processes. This is our focus here, as
opposed to squark and gluino pair production. The sleptons can decay
via R-parity violating operators
\begin{align}
\ssnu_i &\rightarrow 
\left\{
\begin{array}{lll}
\ell_j^+\ell_k^-,\quad &L_iL_j\bar E_k\,,\quad&\mathrm{(a)}
\\[2mm]
d_j \bar d_k,\quad &L_iQ_j\bar D_k\,,\quad&\mathrm{(b)}
\end{array}\right. \label{Eq:decay1}
\end{align}
\begin{align}
\slepton_i^-&\rightarrow
\left\{
\begin{array}{lll}
\bar\nu_j\ell_k^-,\quad &L_iL_j\bar E_k\,,\quad &\mathrm{(a)}
\\[2mm]
\bar u_j d_k,\quad& L_iQ_j\bar D_k\,.\quad &\mathrm{(b)} \\
\end{array}\right. \label{Eq:decay2}
\end{align}
The sleptons can also decay to neutralinos and charginos
\begin{align}
\ssnu_i &\rightarrow 
\left\{
\begin{array}{ll}
\nu_i \chi^0_j\,,\quad &\mathrm{(a)} \\[2mm]
\ell_i^-\chi^+_j\,,\quad &\mathrm{(b)}
\end{array}\right. \label{Eq:decay3}
\\[2mm]
\slepton_i^-&\rightarrow\left\{
\begin{array}{ll}
\ell_i^- \chi^0_j\,,\quad &\mathrm{(a)} \\[2mm]
\nu_i\chi^-_j\,,\quad &\mathrm{(b)}
\end{array}\right.\,.\label{Eq:decay4}
\end{align}
It is the purpose of this paper to investigate resonant slepton
production at the LHC via an operator $LQ\bar D$. We first consider
the decays via the same operator, resulting in resonant dijet
production. We go beyond previous work by comparing with the
\texttt{ATLAS}~\cite{Aad:2011fq} and \texttt{CMS}~\cite{Chatrchyan:2011ns} data, and thus setting relevant bounds on the
underlying R-parity violating supersymmetric model. 

We then consider the decay of the slepton to a neutralino. As 
we show below this can lead to like--sign dileptons in the final state,
due to the Majorana nature of the neutralinos. We then focus on the
case of muons and compare to the \texttt{ATLAS} like--sign dimuon
search~\cite{ATLAS-CONF-2011-126}.

The phenomenology of resonant slepton production was first studied in~\cite{Dimopoulos:1988fr,Hewett:1998fu,Dreiner:1998gz}. A detailed
discussion focusing on the supersymmetric gauge decays resulting in a
like-sign dilepton signature was presented in~\cite{Dreiner:2000vf,Richardson:2000nt,Dreiner:2000qf,Deliot:2000mf}.
Specific benchmark points were investigated in~\cite{Dreiner:2006sv}.
A trilepton signature via the chargino mode in Eq.~(\ref{Eq:decay3})
was discussed in~\cite{Moreau:1999bt,Moreau:2000bs}. Since then
various aspects have been investigated. Single (squark and) slepton
production leading to single top quark production was discussed in~\cite{Oakes:1997zg,Bernhardt:2008mz}. Resonant slepton production with
a 4th family was discussed in~\cite{Cakir:2011dx}, with an ultra light
gravitino in~\cite{Allanach:2003wz}. All but the latter assumed a
neutralino LSP. Resonant slepton production was also considered in the
context of a $\tilde\tau$-LSP in Ref.~\cite{Dreiner:2008rv}. Resonant
squark and slepton production were suggested as an explanation of the
\texttt{CDF} $Wjj$ anomaly in Ref.~\cite{Kilic:2011sr}.

Resonant slepton production has been directly searched for at the
Tevatron by the \texttt{D\O}~\cite{Abazov:2006ii,Autermann:2006iu,Abazov:2007zz} and \texttt{CDF}
experiments~\cite{Abulencia:2006xm,Aaltonen:2010fv,Abulencia:2005nf,Acosta:2005ij},
setting bounds on the relevant parameters. \texttt{D\O}~\cite{Abazov:2006ii,Autermann:2006iu} focused on the resonant 
production and decay of smuons ($\ssmu$) and muon-sneutrinos
($\ssnu_\mu$) via $\lamp_{211}$. The results were presented as upper
limits on $\lamp_{211}$ in the $(\neut_1,\ssmu)$ mass plane within the
minimal supergravity (mSUGRA) / CMSSM~\cite{Chamseddine:1982jx,Barbieri:1982eh,Hall:1983iz,Kane:1993td} framework. The limits are roughly $\lamp_{211} < 0.04~(0.2)$ for smuon masses $m_{\ssmu}\lesssim200-300~(550)\gev$. As we will see,
our study of the LHC data greatly improves these limits.

\texttt{CDF}~\cite{Abulencia:2007mp} and 
\texttt{D\O}~\cite{Abazov:2006nw} also searched for R-parity violation 
assuming the (R-parity conserving) pair production of neutralinos
and/or charginos. Furthermore, \texttt{CDF} investigated R-parity
violation in stop pair production~\cite{Brigliadori:2008vf}.
Implications on R-parity violating models from R-parity conserving
SUSY searches at the Tevatron have been studied
in~\cite{Chakrabarti:2004yq,Das:2005mr,Dreiner:2011xa}.

The $LQ\bar D$ operator could also lead to resonant squark production
at HERA~\cite{Butterworth:1992tc}. This has been searched for by both
\texttt{H1}~\cite{Aktas:2004ij} and \texttt{ZEUS}~\cite{zeus:2006je}.
They obtain limits in terms of a squark mass. For example for a R-parity violating coupling of electromagnetic strength, $\lam'_{11k}=0.3~(k\in\{1,2\})$, the mass bound on the corresponding right-handed down-type squark is $m_{\tilde d_k}\gtrsim 280\gev$~\cite{Aktas:2004ij}.

There are also a few dedicated searches for R-parity violation at
the LHC. The \texttt{ATLAS} collaboration has searched for resonant
tau sneutrino ($\tilde\nu_\tau$) production followed by the R-parity
violating decay to an $e\mu$ final state, \textit{cf.}
Eq.~(\ref{Eq:decay1}a)~\cite{Collaboration:2011qr}. Furthermore, a search for displaced vertices arising from R-parity violating decays of a long-lived neutralino has been performed by \texttt{ATLAS}~\cite{Aad:2011zb}. The \texttt{CMS}
collaboration has considered hadronic supersymmetric pair production
followed by cascade decays to a neutralino. The neutralino then decays
to a purely leptonic final
state~\cite{Chatrchyan:2011ff,CMS-PAS-EXO-11-045}. The
\texttt{ATLAS} collaboration has furthermore interpreted a generic 
search in terms of bounds on a bilinear R-parity violating model
\cite{ATLAS:2011ad}. These are models where $\lam_{ijk},
\;\lam'_{ijk}=0$ and $\kappa_i\not=0$, \textit{cf.} 
Eq.~(\ref{Eq:WnPsix}). In general at any given energy scale $\kappa_i$
can be rotated to zero \cite{Hall:1983id,Banks:1995by}, and we prefer
to work in this basis.

The combined mass limits from LEP, assuming the R-parity violating
decay of pair-produced gauginos or sleptons via $LQ\bar D$ couplings, are
$m_{\neut_1} \ge 39\gev$, $m_{\charge_1^\pm} \ge 103\gev$,
$m_{\ssnu_{\mu,\tau}} \ge 78\gev$ and $m_{\ssmu} \ge
90\gev$~\cite{Heister:2002jc,Barbier:2004ez}. Note however, that the gaugino mass limits are formally only valid in the supersymmetric parameter region investigated by LEP, \ie for a ratio of the Higgs vacuum expectation values of $1\le \tanb \le 35$, a universal soft-breaking scalar mass parameter $m_0 \le 500\gev$, a Higgs mixing parameter $|\mu|\le 200\gev$, a $SU(2)$ gaugino mass parameter $M_2\le 500\gev$ and a R-parity violating coupling larger than $10^{-4}$.

Upper bounds on single $LQ\bar D$ couplings from flavor 
physics and/or from atomic parity violation have been derived and
summarized in \cite{Dreiner:1997uz,Bhattacharyya:1996nj,Allanach:1999ic,Barbier:2004ez,Kao:2009fg}. These bounds usually scale with the
up- or down-type squark mass and thus basically do not constrain
R-parity violating effects in the case where the squarks are decoupled
from the low energy spectrum, which is the case in our analyses.

\section{Resonant sleptons at the LHC}

\subsection{Production process}
\label{Sect:Production}

We consider the single production of a slepton at the LHC,
Eqs.~(\ref{Eq:production1}), (\ref{Eq:production2}). Note that only
the $SU(2)$ doublet left-handed component of the slepton field couples
to this operator. We assume the singly produced slepton to be purely
left-handed. We therefore omit the subscript $L$ in the following.
The case of non-negligible mixing of the weak eigenstates - as usually
relevant for the third generation slepton, the stau - will be briefly
discussed below.

For resonant production, the next-to-leading order (NLO) calculations
including QCD and supersymmetric QCD corrections have been performed
in
Ref.~\cite{Choudhury:2002aua,Yang:2005ts,Chen:2006ep,Dreiner:2006sv}.
They increase the LO cross section at the $14\tev$ LHC by a $K$-factor
of up to $1.35$ for slepton masses less then $2\tev$, while reducing
the uncertainty from the renormalization and factorization scale
dependence\footnote{We checked this by varying the factorization
scale, $\mu_F$, and renormalization scale, $\mu_R$, over the range
$\smass/2 \le \mu_F,~\mu_R \le 2\smass$ for the $7\tev$ cross section
estimate. The deviations from the value obtained at
$\mu_R=\mu_F=\smass$ are less than $3\%$.} to less than
$5\%$~\cite{Dreiner:2006sv}. Further, the authors of
Ref.~\cite{Dreiner:2006sv} have shown that the dependence on the
parton density function (PDF) parametrization is less than $5\%$ by
comparing the cross sections obtained by the
CTEQ6M~\cite{Pumplin:2005rh} and the
MRST04~\cite{Martin:2002dr,Martin:2004ir} fits. We do not expect these
uncertainties to change dramatically for the LHC at a center-of-mass
energy of $\sqrt{s}=7\tev$ and therefore adopt these numbers for this
study.

\begin{figure}
\scalebox{	0.67}{
\hspace{-1.2cm}
\input{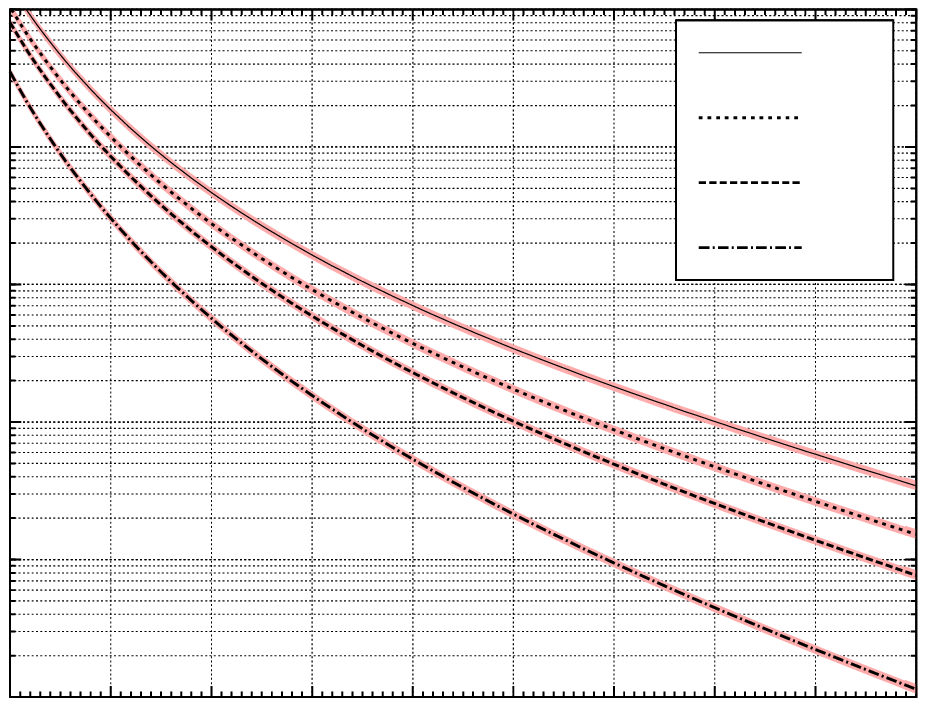}
}
\caption{Single slepton production cross section including QCD NLO 
corrections at the LHC for $\sqrt{s}=7\tev$ as a function of the
slepton mass, $m_{\slepton}$, for ($\lamp_{i11}$, $\lamp_{i12}$,
$\lamp_{i21}$, $\lamp_{i22})=0.01$. The CTEQ6M PDFs have been used,
and renormalization and factorization scales have been identified with
the slepton mass $m_{\slepton}$. The red bands correspond to an
estimated $7\%$ systematic uncertainty including PDF and
renormalization/factorization scale uncertainties.}
\label{Fig:CS}
\end{figure}

The single slepton ($\tilde\nu^{(*)}+\tilde\ell^\pm$) production cross
section at the 7 TeV LHC, including NLO QCD corrections (as employed
here) is shown in Fig.~\ref{Fig:CS}, as a function of the joint slepton
mass, $\smass$. We used the CTEQ6M~\cite{Pumplin:2005rh} PDFs and set
the renormatization and factorization scale equal to the slepton mass,
$\mu_R=\mu_F=\smass$.  The red bands in Fig.~\ref{Fig:CS} indicate the
total theoretical uncertainty of $7\%$, including both scale
uncertainties and PDF parametrization which are added in quadrature.

In Fig.~\ref{Fig:CS} we present the cross sections $\sigma (\lamp, \smass)$ for the R-parity
violating couplings $\lamp_{ijk} = 0.01$ which couple to the first and
second generation quarks ($j,k\in\{1,2\}$). The highest cross section
is obtained for $\lamp_{i11}$ since it involves valence quarks in all
cases. The rate for second generation quarks is suppressed, due to the
lower parton luminosity of the sea quarks. $\sigma(\lamp_{i12})$ is
slightly larger than $\sigma(\lamp_{i21})$ due to the large $u$ quark
flux.

Exemplary event rates are shown in Tab.~\ref{Tab:CSexamples} for $1~\invfb$ of LHC data at $7\tev$. Here
$\lam'=0.01$; the cross section scales with $(\lam')^2$. We further
list the number of singly produced $\slepton^+$, $\slepton^-$ and $
\ssnu+\ssnu^*$ separately. For instance, for a slepton mass $\smass=
500\gev$ and an R-parity violating coupling $\lamp_{i11}=0.01~(0.005)$,
we expect in total $80.8~(20.2)$ signal events, of which the
production of a charged slepton comprises $58\%$. The
$\tilde\ell^+$ rate differs from the $\tilde\ell^-$ rate, since they
involve different parton fluxes. In the case of single stau
production, where the right-handed component of the lightest stau,
$\sstau_1$, cannot be neglected, the cross section is suppressed by
$\cos^2\theta_{\sstau}$, where $\theta_{\sstau}$ is the stau mixing
angle.

\begin{table}
\centering
\renewcommand{\arraystretch}{1.1}\addtolength{\tabcolsep}{0.1cm}
\begin{tabular*}{0.48\textwidth}{@{\extracolsep{\fill}}cccccc}
\hline\hline
$\lamp_{ijk}$ & $\smass$ [GeV] & $\slepton^+$ & $\slepton^-$ & $\ssnu+\ssnu^*$
& total \\
\hline
				& $250$				&	$365$			&	$194$			&	$428$			&	$987$		\\
$\lamp_{i11}=0.01$ 	& $500$				&	$32.6$			&	$14.4$			&	$33.8$			&	$80.8$		\\
				& $800$				&	$4.9$			&	$1.8$			&	$4.5$			&	$11.2$		\\
\hline
				&$250$				&	$275$			&	$47.8$			&	$309$			&	$632$		\\
$\lamp_{i12}=0.01$ 	&$500$				&	$21.8$			&	$2.3$			&	$21.7$			&	$45.8$		\\
				&$800$				&	$2.9$			&	$0.2$			&	$2.6$			&	$5.7$		\\
\hline
				&$250$				&	$40.2$			&	$122$			&	$211$			&	$373$		\\
$\lamp_{i21}=0.01$ 	&$500$				&	$1.8$			&	$7.6$			&	$13.7$			&	$23.1$		\\
				&$800$				&	$0.1$			&	$0.8$			&	$1.6$			&	$2.5$		\\
\hline
				&$250$				&	$25.0$			&	$25.0$			&	$71.5$			&	$122$		\\
$\lamp_{i22}=0.01$ 	&$500$				&	$1.1$			&	$1.1$			&	$3.3$			&	$5.5$		\\
				&$800$				&	$0.06$			&	$0.06$			&	$0.3$			&	$0.42$		\\
\hline\hline
\end{tabular*}
\caption{Number of single slepton events for an integrated luminosity 
of $1\invfb$ at $\sqrt{s}= 7\tev$ using the QCD NLO cross section. The
first column shows the relevant $L_iQ_j\bar D_k$ coupling.  The second
column gives the slepton mass, $\smass$. The third, fourth and fifth
column contain the number of $\slepton^+$, $\slepton^ -$ and
$\ssnu+\ssnu^*$ events. The last column shows the sum.}
\label{Tab:CSexamples}
\end{table}

Although SUSY-QCD corrections can be large in specific regions of the
supersymmetric parameter space~\cite{Dreiner:2006sv}, we do not
include them in order to stay as model-independent as possible.
Next-to-NLO (NNLO) QCD corrections \cite{Majhi:2010zg}, increase the
LHC cross section by 3.4-4\% compared to the NLO result. We do
not include the gluon-gluon fusion production process for sneutrinos,
which is only relevant for $\lamp_{i33}$ \cite{Chen:2006ep}.

\subsection{Slepton decay and signatures}
\label{Sect:decays}

We consider three possible decays of the sleptons. We first analyze the
R-parity violating decay to two jets via the production operator,
\textit{cf.} Eqs.~(\ref{Eq:decay1}b), (\ref{Eq:decay2}b).  The
signature is a narrow dijet resonance. We then consider the decay via
a neutralino or a chargino, \textit{cf.}  Eqs.~(\ref{Eq:decay3}), (\ref{Eq:decay4}). This can lead
to a like-sign dilepton final state signature. For both analyses, we
shall compare our results directly with the relevant
\texttt{ATLAS}~\cite{Aad:2011fq,ATLAS-CONF-2011-126} and
\texttt{CMS}~\cite{Chatrchyan:2011ns} data.

Since we can not perform a detailed analysis while scanning over the
entire supersymmetric parameter space we restrict ourselves to three
specific (simplified) lightest neutralino scenarios:

\begin{enumerate}

\item[\textit{S1}] \textit{bino-like $\neut_1$}: The wino mass $M_2$ 
and the Higgs mixing parameter $\mu$ are much larger than the bino and
the slepton mass ($M_2, \mu \gg M_1, \smass$). $\neut_1$ therefore has
a large bino component. The masses of $\neut_{2,3,4}$, and
$\charge_{1,2}^\pm$, are much larger than $m_{\neut_1}$, and $\smass$.

\item[\textit{S2}] \textit{wino-like $\neut_1$}: $M_1,\mu\gg M_2,
\smass$. Here, $\neut_1$ has a large wino component and it is nearly 
mass degenerate with the (wino-like) $\charge_1^\pm$. $\neut_{2,3,4}$
and $\charge_2^\pm$ are again decoupled from the relevant mass
spectrum.

\item[\textit{S3}] \textit{higgsino-like $\neut_1$}: $M_1,M_2\gg
\mu,\smass$. Here, $\neut_{1,2}$ and
 $\charge_1^\pm$ are nearly mass degenerate and
have a large higgsino component. Hence, gauge interactions of these
sparticles are suppressed. The heavier neutralinos, $\neut_{3,4}$, and
the heavy chargino, $\charge_2^\pm$, are decoupled from the relevant
mass spectrum.
\end{enumerate}
Note that all model parameters in this study are defined at the weak scale.

Within the framework of the CMSSM, the lightest
neutralino is typically dominated by its bino component. Thus, our
first simplified scenario \textit{S1} can be seen as
a good approximation to wide regions of the CMSSM, where the
resonantly produced slepton is lighter than the wino-like $\neut_2$
and $\charge_1^\pm$. In contrast, in anomaly mediated SUSY breaking
scenarios (AMSB)~\cite{Randall:1998uk,Giudice:1998xp,Bagger:1999rd,Baer:2010uy} the lightest neutralino is rather wino-like. For
these scenarios our simplified model \textit{S2} can be considered as
an approximation. Note that this discussion neglects the influence of
the Higgs mixing parameter $\mu$.  In the case of a very small value
of $\mu$ the $\neut_1$ becomes higgsino-like and thus the scenario
takes on the properties of our simplified model \textit{S3}. See also
\cite{Choi:2007ka}.

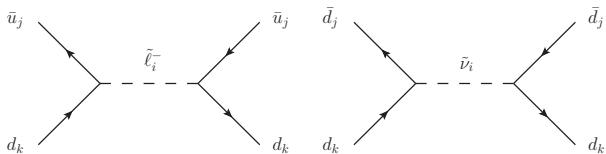
\begin{figure}
\centering
\scalebox{0.48}{
\fcolorbox{white}{white}{
  \begin{picture}(230,170) (100,20)
    \SetWidth{1.0}
    \SetColor{Black}
    \Line[arrow,arrowpos=0.5,arrowlength=5,arrowwidth=2,arrowinset=0.2](180,86)(132,134)
    \Line[arrow,arrowpos=0.5,arrowlength=5,arrowwidth=2,arrowinset=0.2](132,38)(180,86)
    \Line[dash,dashsize=8](180,86)(256,86)
    \Line[arrow,arrowpos=0.5,arrowlength=5,arrowwidth=2,arrowinset=0.2](304,134)(256,86)
    \Line[arrow,arrowpos=0.5,arrowlength=5,arrowwidth=2,arrowinset=0.2](256,86)(304,38)
    \Text(108,130)[lb]{\Large{\Black{$\bar u_j$}}}
    \Text(108,32)[lb]{\Large{\Black{$d_k$}}}
    \Text(215,96)[lb]{\Large{\Black{$\slepton_{i}^-$}}}
    \Text(315,130)[lb]{\Large{\Black{$\bar u_j$}}}
    \Text(315,32)[lb]{\Large{\Black{$d_k$}}}    
  \end{picture}
  }
\fcolorbox{white}{white}{
  \begin{picture}(230,170) (100,20)
    \SetWidth{1.0}
    \SetColor{Black}
    \Line[arrow,arrowpos=0.5,arrowlength=5,arrowwidth=2,arrowinset=0.2](180,86)(132,134)
    \Line[arrow,arrowpos=0.5,arrowlength=5,arrowwidth=2,arrowinset=0.2](132,38)(180,86)
    \Line[dash,dashsize=8](180,86)(256,86)
    \Line[arrow,arrowpos=0.5,arrowlength=5,arrowwidth=2,arrowinset=0.2](304,134)(256,86)
    \Line[arrow,arrowpos=0.5,arrowlength=5,arrowwidth=2,arrowinset=0.2](256,86)(304,38)
    \Text(108,130)[lb]{\Large{\Black{$\bar d_j$}}}
    \Text(108,32)[lb]{\Large{\Black{$d_k$}}}
    \Text(215,96)[lb]{\Large{\Black{$\ssnu_{i}$}}}
    \Text(315,130)[lb]{\Large{\Black{$\bar d_j$}}}
    \Text(315,32)[lb]{\Large{\Black{$d_k$}}}    
  \end{picture}  
}}
\caption{Resonant production of a charged slepton, $\slepton_{i}^-$, 
(left) and a sneutrino, $\ssnu_i$, (right), followed by the direct 
decay into two quarks via the R-parity violating coupling $\lamp_
{ijk}$. This process leads to a narrow dijet resonance.}
\label{Fig:dijetprocess}
\end{figure}

The resonant dijet processes via the operator $\lam'_{ijk}$,
Eqs.~(\ref{Eq:production1}), (\ref{Eq:production2}) and the decays
Eqs.~(\ref{Eq:decay1}), (\ref{Eq:decay2}), are depicted in
Fig.~\ref{Fig:dijetprocess}. At tree level, the decay width is $\Gamma
(\slepton_i^- \to \bar u_j d_k)\approx75\,$MeV, for $\smass = 500\gev$
and $\lamp=0.05$ \cite{Richardson:2000nt}.  At hadron colliders, this
process leads to a very narrow resonance in the invariant mass
spectrum of the dijet system. However, due to the large QCD background
at the LHC it will only be visible for large slepton masses $\tilde m
\gtrsim 1\tev$ and reasonably large R-parity violating couplings
$\lamp \gtrsim \mathcal{O}(10^{-2})$.

If the slepton or sneutrino is the LSP, the dijet channel is the only
kinematically allowed decay mode. For a $\neut_1$ LSP, the slepton
decay to dijets is competing with the R-parity conserving decay
$(\slepton/\ssnu)\to (\ell / \nu)+\neut_1$ and possibly other decays
to lighter sparticles, \cf Eqs.~\eqref{Eq:decay3}, \eqref{Eq:decay4}.
A typical value for the kinematically unsuppressed ($m_{\neut_1} \ll
\smass$) decay width is $\Gamma(\slepton\to\ell\neut_i)\approx1\,
$GeV, for $\tilde m=500\,$GeV \cite{Haber:1984rc,Dreiner:2008tw}. This
broadens the dijet resonance, and reduces the dijet branching
ratio. The exact branching ratios depend on the R-parity violating
coupling strength $\lamp$, the composition of the light gauginos and
on the details of the mass spectrum. The gauge decays are basically
absent in \textit{S3} for the first and second generation sleptons,
but can be relevant for a scalar tau.

If $\chi^0_1$ is the LSP it decays via the operator $L_iQ_j\bar D_k$ as
\begin{eqnarray}
\chi^0_1\rightarrow\left\{ \begin{array}{l}
\ell^-_iu_j\bar d_k \\[2mm]
\nu_i d_j\bar d_k 
\end{array} +\mathrm{c.c}.
\right. \label{Eq:neutralino-decay}
\end{eqnarray}
The complex conjugate decays are equally likely, due to the Majorana
nature of the neutralino. The neutrino and charged lepton decay modes
can have different branching ratios depending on the admixture of the
lightest neutralino. The decay $\chi^0_1\rightarrow\nu_i
\gamma$ for $L_iQ_j\bar D_k$ is only possible for $j=k$ 
\cite{Hall:1983id} but is typically highly suppressed and
not relevant for collider signatures \cite{Dreiner:1991pe}. 

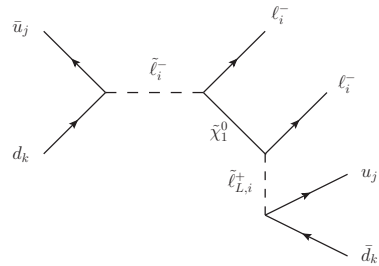
\begin{figure}
\centering
\scalebox{0.48}{
\fcolorbox{white}{white}{
  \begin{picture}(300,210) (100,-50)
    \SetWidth{1.0}
    \SetColor{Black}
    \Line[arrow,arrowpos=0.5,arrowlength=5,arrowwidth=2,arrowinset=0.2](180,86)(132,134)
    \Line[arrow,arrowpos=0.5,arrowlength=5,arrowwidth=2,arrowinset=0.2](132,38)(180,86)
    \Line[dash,dashsize=8](180,86)(256,86)
    \Line[arrow,arrowpos=0.5,arrowlength=5,arrowwidth=2,arrowinset=0.2](256,86)(304,134)
    \Line[](304,38)(256,86)
    \Line[dash,dashsize=8](304,38)(304,-10)
    \Line[arrow,arrowpos=0.5,arrowlength=5,arrowwidth=2,arrowinset=0.2](304,38)(352,86)
    \Line[arrow,arrowpos=0.5,arrowlength=5,arrowwidth=2,arrowinset=0.2](304,-10)(368,22)
    \Line[arrow,arrowpos=0.5,arrowlength=5,arrowwidth=2,arrowinset=0.2](368,-42)(304,-10)
    \Text(108,130)[lb]{\Large{\Black{$\bar u_j$}}}
    \Text(108,32)[lb]{\Large{\Black{$d_k$}}}
    \Text(215,96)[lb]{\Large{\Black{$\slepton_{i}^-$}}}
    \Text(310,140)[lb]{\Large{\Black{$\ell_i^-$}}}
    \Text(262,48)[lb]{\Large{\Black{$\neut_1$}}}
    \Text(362,86)[lb]{\Large{\Black{$\ell_i^-$}}}
    \Text(276,6)[lb]{\Large{\Black{$\slepton_{L,i}^+$}}}
    \Text(380,16)[lb]{\Large{\Black{$u_j$}}}
    \Text(380,-48)[lb]{\Large{\Black{$\bar d_k$}}}
  \end{picture}
}}
\caption{Resonant production of a charged slepton, $\slepton_{i}$, 
with successive decay into the lightest neutralino, $\neut_1$, and 
a charged lepton $\ell_i$. The subsequent decay of the $\neut_1$ 
can lead to another lepton of the same charge due to the Majorana 
nature of the neutralino. Thus, this process gives rise to a like--sign 
dilepton signature.}
\label{Fig:dileptonprocess}
\end{figure}

Within the framework of the three decoupled scenarios
\textit{S1}-\textit{S3}, only the process
\begin{align}
\bar u_j d_k \to \slepton^- \to \ell^- \neut_{1}\qquad\qquad\nonumber\\
\qquad\qquad\overset{\lamp}{\hookrightarrow} \ell^- u_j \bar d_k~ 
\label{Eq:dilepton_sig1}
\end{align}
(and its charged conjugate), can lead to a like--sign dilepton
signature. One diagram contributing to this process is also
illustrated in Fig.~\ref{Fig:dileptonprocess}. The sneutrino
production
\begin{equation}
d_j \bar d_k \to \ssnu^* \to \ell^+ \charge_1^-
\end{equation}
followed by the decay of the chargino $\tilde \chi^-_1\rightarrow
\ell^-_i\bar d_j d_k $ (via $L_iQ_j\bar D_k$) leads to an
opposite-sign dilepton signature. The cascade decay of the chargino
via the neutralino
\begin{align}
d_j \bar d_k \to \ssnu^* \to \ell^+ \charge_1^-\qquad\qquad\quad\;\qquad\nonumber\\
\qquad\qquad\hookrightarrow W^- \neut_1\qquad\qquad\nonumber\\
\qquad\qquad\overset{\lamp}{\hookrightarrow} \ell^+ \bar u_j d_k
\end{align}
in the wino-like scenario is kinematically suppressed since $\charge_1
^-$ and $\neut_1$ are nearly mass degenerate. 

In {\it S3}, $\neut_1$ can be replaced by $\neut_2$ in
Eq.~\eqref{Eq:dilepton_sig1}. The $\neut_1$ and $\neut_2$ have similar
couplings due to their large higgsino component and are again nearly
mass degenerate. Thus, this process contributes with a similar rate to
the like--sign dilepton signature as the process in
Eq.~\eqref{Eq:dilepton_sig1}. In addition, the rate is enhanced by
roughly a factor of 2 compared to the bino- and wino-like $\neut_1$
scenario because the neutral decay $\neut_{1,2}\to \nu_i
d_j\bar d_k$ in Eq.~\eqref{Eq:neutralino-decay} is suppressed for a higgsino $\neut_{1,2}$.

\begin{figure*}
\centering
\scalebox{	0.67}{
\input{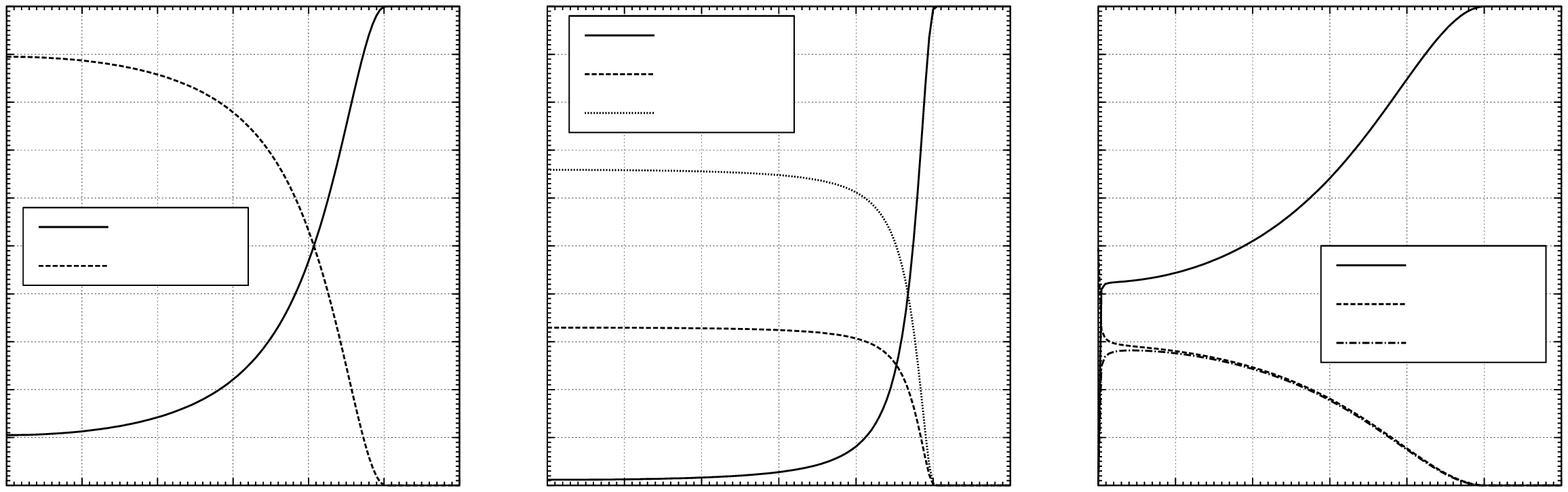}
}
\caption{Neutralino mass dependence of the branching ratios of the 
slepton decay modes in the bino-like (left panel), wino-like
(middle) and higgsino-like (right) $\neut_1$ scenario. We chose a
coupling strength of $\lamp=0.05$. The slepton mass is set to
$\smass=500\gev$. The decays are calculated with {\tt
ISAJET7.64}~\cite{Paige:2003mg}. In the bino- and wino-like $\neut_1$ scenario, the
(purely left-handed) slepton can be $\slepton = \sse_L, \ssmu_L,\sstau_L$, while in the higgsino-like
$\neut_1$ scenario we only show the decays of a (purely left-handed)
$\sstau_L$. We set $\tan\beta =10$.}
\label{Fig:BRs}
\end{figure*}

\begin{figure*}
\centering
\scalebox{	0.67}{
\input{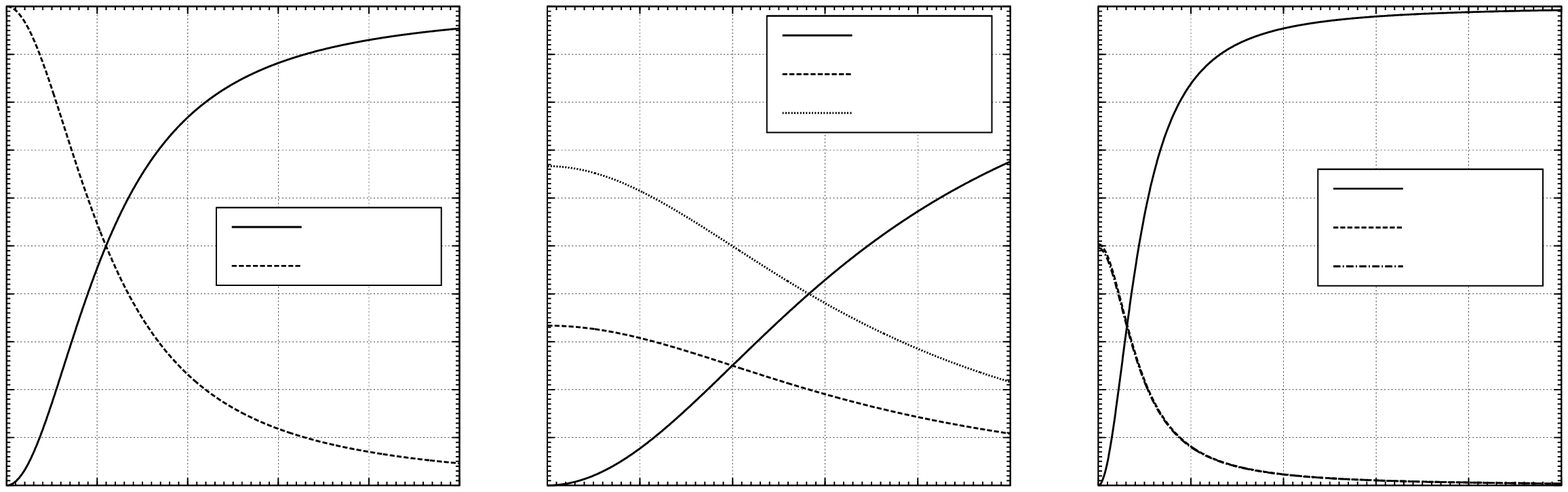}
}
\caption{$\lamp$ dependence of the branching ratios of the slepton 
decay modes in the bino-like (left panel), wino-like (middle)
and higgsino-like (right) $\neut_1$ scenario. We chose a slepton mass of
$\smass=500\gev$ and a lightest neutralino mass of
$m_{\neut_1}=250\gev$. The decays are obtained with {\tt
ISAJET7.64}~\cite{Paige:2003mg}. As in Fig.~\ref{Fig:BRs} the (purely left-handed) slepton
can be $\slepton = \sse_L, \ssmu_L,
\sstau_L$, in the bino- and wino-like $\neut_1$ scenario, while in the
higgsino-like $\neut_1$ scenario we only show the decays of a (purely
left-handed) $\sstau_L$. We set $\tan\beta =10$.}
\label{Fig:BR_lamp}
\end{figure*}

In Fig.~\ref{Fig:BRs} and ~\ref{Fig:BR_lamp} we show the dependence of
the charged slepton branching ratios corresponding to the decays
Eq.~(\ref{Eq:decay2}b) and \eqref{Eq:decay4}, on the
lightest neutralino mass, $m_{\neut_1}$, and coupling strength,
$\lamp$, respectively.  In both figures we chose a slepton mass of
$\smass = 500\gev$. In Fig.~\ref{Fig:BRs} the R-parity violating
coupling strength is set to $\lamp =0.05$. In Fig.~\ref{Fig:BR_lamp}
we fixed the lightest neutralino mass to $250\gev$.

As $m_{\neut_1}$ increases\footnote{Computationally, we increase
$M_1$, $M_2$ or $\mu$ in the bino-, wino- or higgsino-like $\neut_1$
scenario, respectively, while setting the decoupled mass
parameters to $5\tev$.}, the phase space in the gauge decays of the
slepton, Eq.~\eqref{Eq:decay4}, decreases and leads to a suppression
of the R-parity conserving decays, Eq.~(\ref{Eq:decay2}b). For
$m_{\neut_1}\ge 500\gev$, the slepton becomes the LSP and only the
dijet decay channel remains accessible. Note that there are extensive
regions in R-parity violating CMSSM parameter space where the slepton
is indeed the LSP
\cite{Dreiner:2008rv,Bernhardt:2008jz,Dreiner:2008ca,Desch:2010gi,Dreiner:2011xa}.

For the bino- and wino-like $\neut_1$ scenario (the left and middle
panels in Fig.~\ref{Fig:BRs} and \ref{Fig:BR_lamp}, respectively), we
show the branching ratios $\mathcal{B}(\slepton^+ \to u \bar d)$,
$\mathcal{B}(\slepton \to \ell \neut_{1})$ and
$\mathcal{B}(\slepton^+ \to \bar\nu \charge_1^+)$, where the charged
slepton is the left-handed slepton of any of the three generations,
$\slepton = \sse_L, \ssmu_L, \sstau_L$.

In {\it S1}, the only kinematically allowed slepton decays are
$\slepton^+\to u \bar d$ and $\slepton\to \ell \neut_1$. Recall
$M_2$ is very large and thus the lightest chargino is heavy. {\it S1}
can be viewed as the best-case scenario for the like--sign dilepton
signature because the gauge decay of the charged slepton leads in
roughly $25\%$ of the cases to the like--sign dilepton signature. The
decay $\slepton \to\ell \neut_1$ dominates for $\lamp\lesssim
0.05~(0.1)$ given a sufficiently large phase space of $\smass -
m_{\neut_1}\gtrsim100~(250)\gev$.

In the wino-like $\neut_1$ scenario, we have the three competing decays
$\slepton^+ \to u \bar d$, $\slepton \to \ell \neut_1$ and 
$\slepton^+ \to \bar\nu \charge_1^+$. The slepton decays twice as
often to the chargino as to the neutralino, $\mathcal{B}(\slepton^+\to\bar\nu
\charge_1^+) \approx 2 \mathcal{B}(\slepton \to \ell \neut_1)$. The 
gauge decays of the charged slepton therefore yield a like--sign
dilepton signature only around 1/12 of the time. The gauge decays
dominate for $\lamp \lesssim 0.05~(0.35)$ for a mass difference of
$m_{\slepton} - m_{\neut_1} \gtrsim 50~(250)\gev$. They are slightly
stronger than in the bino-like $\neut_1$ case due to the larger gauge
coupling.

In the higgsino-like $\neut_1$ scenario (the right panel in
Fig.~\ref{Fig:BRs} and \ref{Fig:BR_lamp}), we only give the branching
ratios of the (left-handed\footnote{Here, we decoupled the soft-breaking right-handed stau mass parameter, $(m_{\tilde{\mathbf{E}}})_{33} = 5\tev$, which leads to the lightest stau being purely left-handed.}) third generation slepton, $\sstau_L$,
because of the non-negligible Higgs Yukawa couplings. The gauge decays
of the first and second generation sleptons are
negligible. These thus only decay to dijets.

We therefore discuss the higgsino-like $\neut_1$ scenario only for a
left-handed $\sstau_1$. For this, we set the ratio of the Higgs vacuum
expectation values, $\tan\beta = 10$, which influences the $\tau$
Yukawa coupling. The branching ratios $\mathcal{B}(\sstau\to\tau\neut
_{1,2})$ are roughly equal. The gauge decays of the stau yield
a like--sign tau pair $50\%$ of the time.  However, they
dominate the slepton decay modes only for a coupling $\lamp \lesssim
0.01~(0.04)$ for a given mass difference of $m_{\slepton} -
m_{\neut_1} \gtrsim 50~(250)\gev$.

In the case of the lightest stau, $\sstau_1$, having a non-negligible
right-handed component the branching ratios get more complicated. The
right-handed component does not couple to the R-parity violating
operator but via Yukawa interactions to the chargino, leading to the
decay $\sstau_R^+ \to \bar\nu_\tau \charge_1^+$. Therefore, with
increasing right-handedness of the $\sstau_1$, the R-parity violating
decay mode to two jets on the one hand gets suppressed while on the
other hand the additional decay mode to the chargino decreases the
(like-sign) dilepton rate. Recall that the production is
also suppressed for a right-handed stau.

We do not further consider the higgsino-like $\neut_1$ scenario.
However, this analysis and the following results in
Sect.~\ref{Sect:likesign2m} show that a search for like--sign tau
pairs would be able to probe resonantly produced tau sleptons with
$\lamp_{3jk}$ ($j,k=1,2$) even if the light gauginos, $\neut_{1,2}$
and $\charge_1^\pm$, are dominated by their higgsino component.


\section{Searches at the LHC}

In this section we use both the dijet and the like--sign dilepton
signatures of resonant slepton production to constrain the R-parity
violating couplings $\lamp_{ijk}$ and the relevant slepton
mass. For the calculation of both the R-parity conserving and
violating sparticle decays we use {\tt ISAJET7.64}~\cite{Paige:2003mg}
and {\tt ISAWIG1.200}~\cite{ISAWIG}. The {\tt ISAWIG} output is fed
into {\tt
Herwig6.510}~\cite{Corcella:2000bw,Corcella:2002jc,Moretti:2002eu} for
the MC simulation at particle level. We simulate the response of the
\texttt{ATLAS} and \texttt{CMS} detector using the general purpose
detector simulation package {\tt Delphes1.9}~\cite{Ovyn:2009tx}. Jets
are reconstructed using the anti-$k_T$
algorithm~\cite{Cacciari:2008gp,Cacciari:2005hq}. In the dijet
resonance search in Sect.~\ref{Sect:2j}, the distance parameter is set
to $R=0.6$ (\texttt{ATLAS}) and $R=0.5$ (\texttt{CMS}), while we use
$R=0.4$ for the \texttt{ATLAS} like--sign dimuon search in
Sect.~\ref{Sect:likesign2m}. These jet definitions are in accordance
with Refs.~\cite{Aad:2011fq,Chatrchyan:2011ns,ATLAS-CONF-2011-126}.


\subsection{Search for dijet resonances}
\label{Sect:2j}

Both the \texttt{ATLAS}~\cite{Aad:2011fq} and the
\texttt{CMS}~\cite{Chatrchyan:2011ns} experiment have searched for resonances
in the dijet invariant mass spectrum using $pp$ collision data
corresponding to an integrated luminosity of $1.0\invfb$ at a
center-of-mass energy of $\sqrt{s}=7\tev$. The non-observation of new
resonances led the experiments to derive limits for several new
physics models including string resonances, exited quarks, axigluons
and colour octet scalar resonances.  In the following, we use the
model-independent limits on a fiducial signal cross section provided
by \texttt{ATLAS}~\cite{Aad:2011fq} and
\texttt{CMS}~\cite{Chatrchyan:2011ns} to constrain the
resonant R-parity violating production of sleptons,
Eqs.~(\ref{Eq:production1}), (\ref{Eq:production2}), with subsequent
decay to two jets, Eqs.~(\ref{Eq:decay1}b), (\ref{Eq:decay2}b). The mass region in the \texttt{ATLAS} (\texttt{CMS}) 
search ranges from $0.9\tev$ ($1\tev$) to $4.0\tev$ ($4.1
\tev$). Therefore, these searches can only constrain the resonant 
production of very heavy sleptons. Constraints for lower
slepton masses have been derived from \texttt{CDF} and \texttt{UA2}
searches in~\cite{Kilic:2011sr}.

In order to evaluate the acceptance of the analyses, we simulated
25,000 signal events for the process $pp \to \slepton_i / \ssnu_i \to
q_j q_k$ for each slepton mass, $\smass$. For the \texttt{ATLAS}
search, we followed closely the prescription given in the Appendix of
Ref.~\cite{Aad:2011fq}. There, the limits are presented assuming a
certain width to mass ratio of the resonance, $\sigma_G/m_G$. In our
study we determined $\sigma_G/m_G$ with Gaussian fits of the dijet
invariant mass distribution in the region between $0.8 \smass$ and
$1.2 \smass$. It ranges from $8\%$ to $5\%$ for slepton masses from
$0.9\gev$ to $4\tev$. The acceptance $\mathcal{A}$ is given by the
fraction of events lying in the region $0.8 \smass$ to $1.2 \smass$
(after all other kinematic requirements are applied) and ranges from
$8.1\%$ to $18.6\%$ for slepton masses from $0.9\tev$ to $4\tev$.

\begin{figure}
\scalebox{	0.67}{
\hspace{-0.4cm}
\input{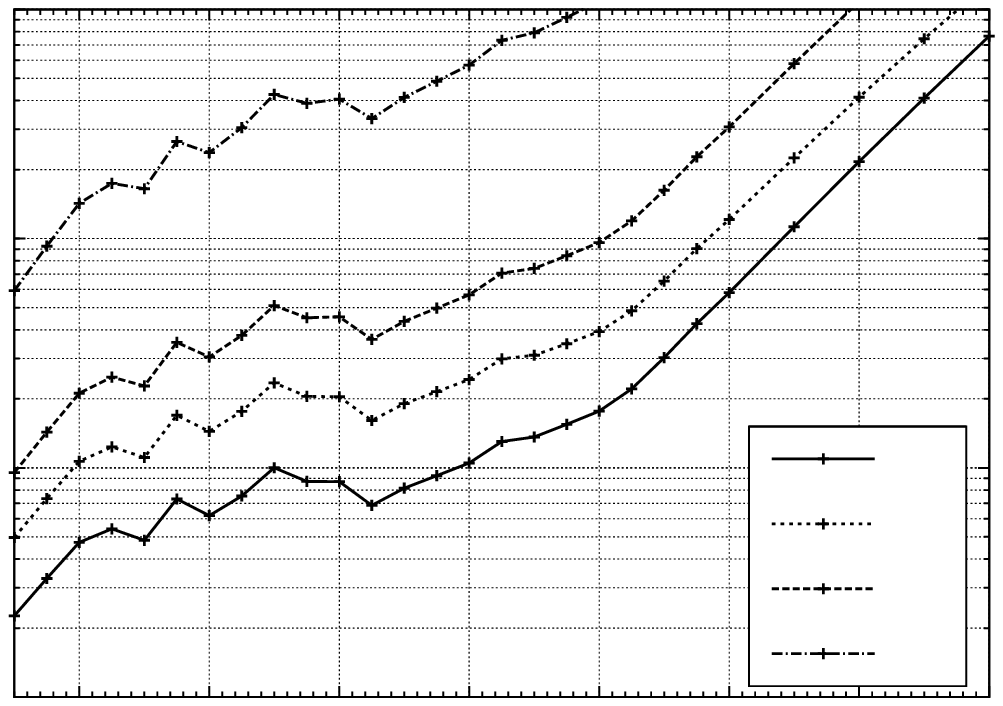}
}
\caption{Upper bounds on ${\lamp}^2 \times \mathcal{B}(\slepton_i/\ssnu
_i\to jj)$ derived from the \texttt{ATLAS} dijet resonance searches 
with $1\invfb$ of data.}
\label{Fig:ATLAS_dijet_results}
\end{figure}

Both $\mathcal{A}$ and $\sigma_G/m_G$ are fairly independent of $\lamp
_{ijk}~(j,k\in\{1,2\})$ for values between 0.001 and $1.0$, since the
resonance shape is dominated by the jet smearing of the detector
simulation. Thus, we can easily derive upper limits on the R-parity
violating coupling squared times the branching ratio to dijets of the
resonant slepton, ${\lamp} ^2\times \mathcal{B}(\slepton_i/\ssnu_i\to
jj)$, for a given resonant slepton mass, $\smass$. These
limits\footnote{This analysis assumes that the sneutrino and
the charged slepton resonance are not distinct. This is generally the
case as long as the mass splitting is not too large, \ie $m_{\slepton}
- m_{\ssnu} \lesssim \sigma_G \lesssim 10\%\,m_{\slepton}$.} are shown
in Fig.~\ref{Fig:ATLAS_dijet_results} for the four types of couplings
$\lamp_{i11}$, $\lamp_{i12}$, $\lamp_{i21}$ and $\lamp_{i22}$
($i=1,2,3$). In the case of an intermediate third generation slepton
($i=3$), the limit has to be multiplied by $\cos^2\theta_{\sstau}$ to
account for possible mixing in the stau sector. To be conservative, we
reduced the signal by $7\%$ to take into account the theoretical
uncertainty of the NLO cross section prediction. The statistical
uncertainty of the acceptance estimate is negligible.

The upper bounds on the four investigated R-parity violating
couplings, as derived from the \texttt{ATLAS} search, are listed
together with $\mathcal{A}$ and $\sigma_G/m_G$ in
Tab.~\ref{Tab:ATLASdijet} in Appendix~\ref{App:add}. We only show
upper limits for values $\lamp < 1$ (perturbativity). For instance,
assuming the decay to dijets being the only accessible decay mode, we
can derive the upper bounds $\lamp_{i11} \le 0.07~(0.09)$ and
$\lamp_{i22} \le 0.38~(0.64)$ for a slepton mass $\smass =
1000\gev~(1500\gev)$.

In the \texttt{CMS} search~\cite{Chatrchyan:2011ns}, so--called wide
jets are constructed based on anti-$k_T$ jets with distance parameter
$R=0.5$. This allows to distinguish between a quark--quark (qq),
quark--gluon (qg) and a gluon--gluon (gg) dijet system. Here, we employ
the 95\% CL upper limits on $\sigma\times\mathcal{A}$ derived for a qq
dijet system. These limits only assume the natural resonance width to
be small compared to the \texttt{CMS} dijet mass resolution.

We adopt the \texttt{CMS} construction of wide jets and apply the
kinematic requirements to the jets. The acceptance is defined by the
fraction of events with dijet invariant mass $m_{jj}>838\gev$. It
ranges from $33.8\%$ to $44.8\%$ for slepton masses from $1.0\tev$ to
$4.1\tev$. Again, we take into account a $7\%$ systematic uncertainty
on the signal.

\begin{figure}
\scalebox{	0.67}{
\hspace{-0.4cm}
\input{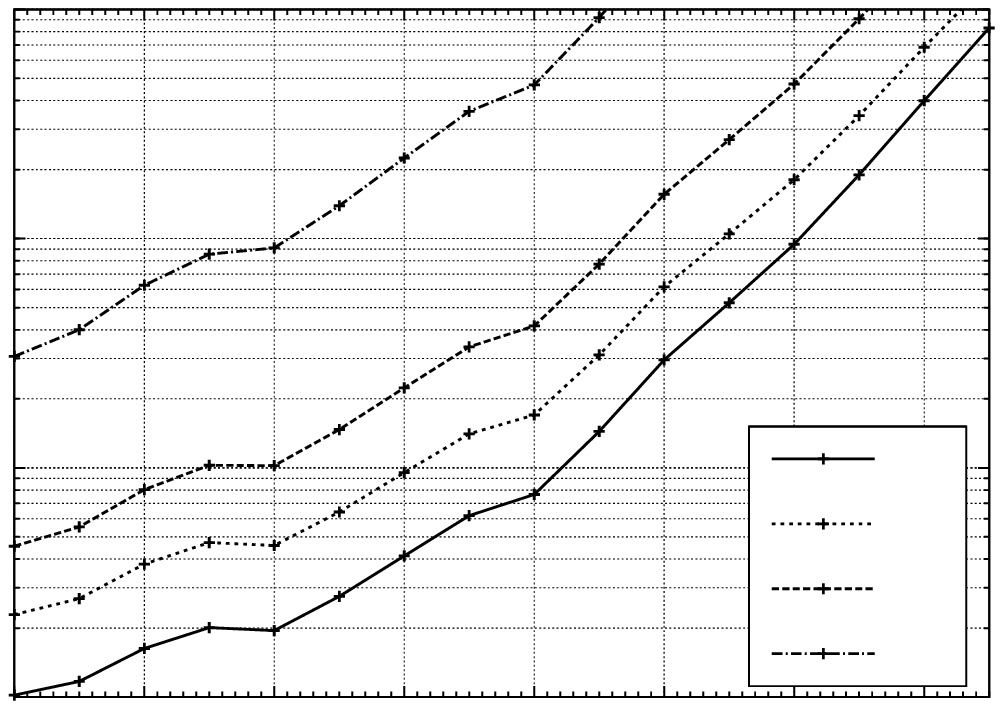}
}
\caption{Upper bounds on ${\lamp}^2 \times \mathcal{B}(\slepton_i / 
\ssnu_i \to jj)$ derived from the \texttt{CMS} dijet resonance 
searches with $1\invfb$ of data.}
\label{Fig:CMS_dijet_results}
\end{figure}

In Fig.~\ref{Fig:CMS_dijet_results} we present the upper
bounds on ${\lamp}^2 \times \mathcal{B}(\slepton_i / \ssnu_i \to jj)$
for the same couplings as before, but now derived from the
\texttt{CMS} search. These results are given in detail in
Tab.~\ref{Tab:CMSdijet} in Appendix~\ref{App:add}. For a pure dijet
decay of the slepton, $\mathcal{B}(\slepton_i/\ssnu_i \to jj) \approx
100\%$, the upper bounds obtained are $\lamp_{i11} \le 0.03~(0.05)$
and $\lamp_{i22} \le 0.18~(0.37)$ for a slepton mass $\smass =
1000\gev~(1500\gev)$. Due to the higher acceptance of the \texttt{CMS}
search, these limits are considerably stricter than those obtained from the
\texttt{ATLAS} search.

\subsection{Search for prompt like--sign dimuons}
\label{Sect:likesign2m}

\begin{figure*}[t]
\subfigure[\ $p_T$ distribution of the isolated muons.]{
\scalebox{	0.645}{
\hspace{-0.4cm}
\input{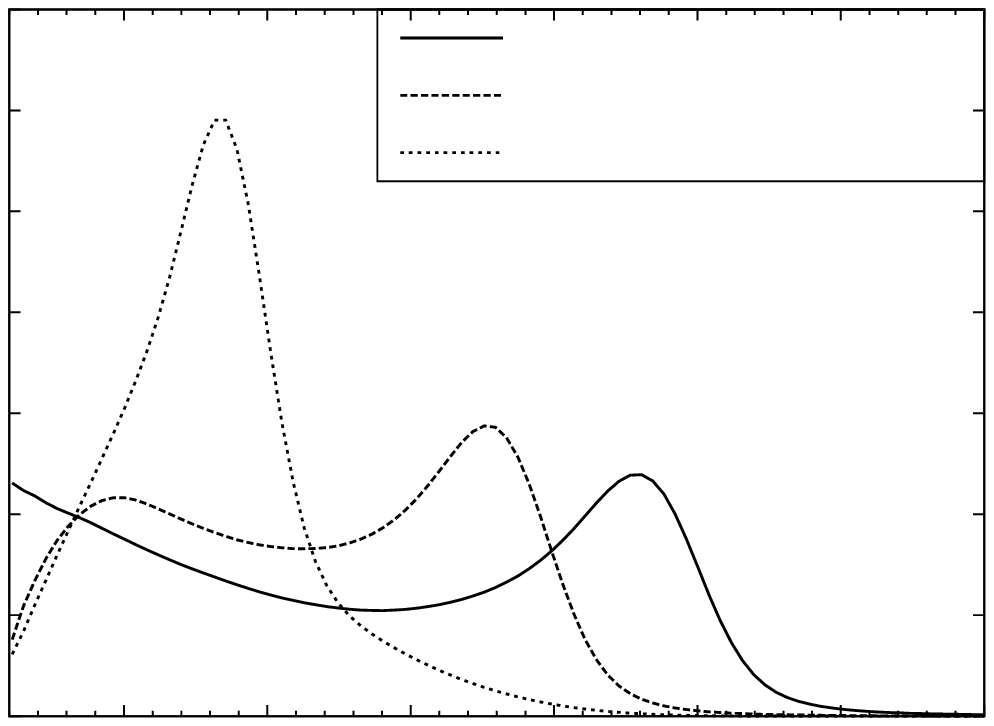}
}
\label{Fig:lepPT}
}
\subfigure[\ Invariant mass distribution of the like--sign dimuon pairs.]{
\scalebox{	0.645}{
\hspace{-0.4cm}
\input{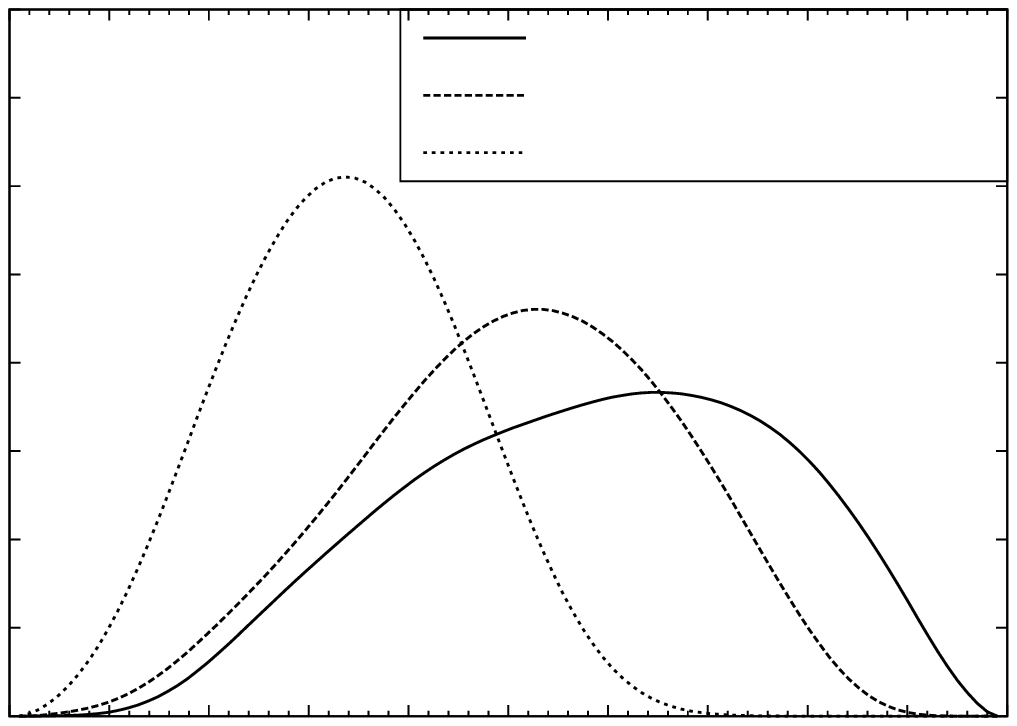}
}
\label{Fig:Mll}
}
\caption{Kinematic properties of the single slepton production process $pp \to \slepton^* / \ssnu^* \to \mu/\nu \neut_1$ via $\lamp_{2jk}$ at the LHC with a center-of-mass energy of $7\tev$: (a)
Transverse momentum distribution of the muons passing the object
selection (isolation, $p_T > 10\gev$) of
Ref.~\cite{ATLAS-CONF-2011-126}; (b) Invariant mass distribution of
the like--sign dimuon pairs which pass the full event selection. The
slepton mass is set to $\smass = 500\gev$. We show the shapes for
three different neutralino masses, $m_{\neut_1}=(100, 250, 500) \gev$.}
\label{Fig:kinematics}
\end{figure*}

We now turn to the discussion of the constraints from the like--sign
dilepton signature. In Ref.~\cite{ATLAS-CONF-2011-126} \texttt{ATLAS}
searched for anomalous production of prompt like--sign muon pairs,
using data corresponding to an integrated luminosity of $1.6\invfb$ at
a center-of-mass energy of $\sqrt{s}=7\tev$. No significant excess was
observed and upper limits on the anomalous production of prompt
like--sign muon pairs were derived. In the following, we use these
results to constrain the R-parity violating couplings $\lamp_{2jk},\;
j,k\in\{1,2\}$, assuming the resonant production of a left-handed
smuon, $\ssmu_L$ via Eq.~(\ref{Eq:production2}), and its
subsequent decay into the lightest neutralino, $\neut_1$, and
a muon via Eq.~(\ref{Eq:decay4}a). The neutralino then decays as in
Eq.~(\ref{Eq:neutralino-decay}) to the lepton with the same sign
charge.

In the \texttt{ATLAS} search~\cite{ATLAS-CONF-2011-126}, the signal
region is subdivided into four. The signal yield is defined by
the number of like--sign muon pairs whose invariant mass,
$m_{\mu\mu}$, is greater than $25\gev$, $100\gev$, $200\gev$ and
$300\gev$, respectively. The main requirements on the muons are the following: The transverse
momentum of the first (second) muon is larger then $20~(10)\gev$. Both
muons are in the central region of the detector with
pseudorapidity $|\eta|<2.5$. They are separated from jets by $\Delta
R(\mu,\mathrm{ jet}) >0.4$, where jets are defined by the anti-$k_t$
algorithm with a distance parameter of $R=0.4$ and minimal transverse
momentum $p_T(\mathrm{jet}) > 7\gev$. The muons have to be prompt
(originating from the primary vertex). This translates in our case
into a requirement on the slepton lifetime to be less then
$\tau<10^{-14}~\mathrm{s}$. Furthermore, we employ the same cone
isolation criteria for the muons as in the
\texttt{ATLAS} note~\cite{ATLAS-CONF-2011-126}.

We now discuss the kinematic properties of single slepton production
at the LHC with $7\tev$ center-of-mass energy. The slepton is forced
to decay into the lightest neutralino, \ie we consider the process\footnote{We must include
the sneutrino production even though it does not lead to like--sign
dileptons. Both production processes are jointly encoded in \texttt{Herwig6.510}.} $pp\to\slepton^*/\ssnu^*\to(\mu/\nu)\neut_1$. In Fig.~\ref{Fig:lepPT} we provide the transverse momentum
($p_T$) distribution of the muons passing the isolation,
pseudorapidity, jet separation and minimal transverse momentum
($p_T\ge 10\gev$) constraints, whereas Fig.~\ref{Fig:Mll} gives the
invariant mass distribution\footnote{Both distributions in
Fig.~\ref{Fig:kinematics} are obtained from Monte-Carlo simulation
using the bino--like $\neut_1$ scenario, normalized to unity for a bin
size of $2\gev$ and then smoothened for better visualization.} of the
like-sign dimuon pairs after the full event selection (except the
final $m_{\mu\mu}$ requirement). We give these distributions for three
example models with different lightest neutralino masses $m_{\neut_1}
= (100, 250, 400)\gev$, slepton mass $\smass = 500\gev$ and a non-zero
R-parity violating coupling $\lamp_{2jk}$.

For large mass splittings between the slepton and the neutralino,
$\Delta m = \smass - m_{\neut_1}$, we can identify two distinct peaks
in the muon-$p_T$ spectrum. In the first model considered ($\smass
=500\gev$, $m_{\neut_1}=100\gev$), we have hard muons with typical
$p_T$ values around $200-250\gev$. These muons origin from the slepton
decay. In constrast, the soft muons accumulating at the low end of
the distribution stem from the three-body decay of the neutralino (and
the chargino in the wino--like scenario).

For larger neutralino masses (second and third model) the phase space
for the muons from the gaugino decay increases on the one hand,
leading to the migration of the left peak in the $p_T$ distribution
towards higher values. On the other hand, the muons from the slepton
decay become softer due to the smaller $\Delta m$.  In the third model
considered ($\smass =500\gev$, $m_{\neut_1}=400\gev$), the peaks
overlap at a $p_T$ value of around $80-90\gev$. For even smaller
$\Delta m$, the muons from the slepton decay will constitute the low
end of the $p_T$ spectrum.

The invariant mass distribution of the like--sign dimuon pairs shown
in Fig.~\ref{Fig:Mll} exhibits a broad peak of approximately gaussian
shape. The peak value increases for larger mass splitting $\Delta m$.

From this discussion, we can already predict that the acceptance of
the \texttt{ATLAS} like--sign dimuon search will decrease for
(\textit{i}) small neutralino masses and (\textit{ii}) in the small
$\Delta m$ region, where the slepton and the neutralino are close in
mass. In both cases, one of the muons is rather soft due to reduced
phase space and thus may not fulfill the minimum $p_T$
requirement. This is especially important for (\textit{i}) since the
neutralino decays via a three-body decay. On the other hand, in
($\textit{ii}$), the invariant mass $m_{\mu\mu}$ tends to be small,
thus reducing in particular the acceptance of the high $m_{\mu\mu}$
signal regions.

The (normalized) distributions in Fig.~\ref{Fig:kinematics} are to a
good approximation independent of the choice of $j$, $k$ and the value
of $\lamp_{2jk}$ (as long as it is a prompt neutralino decay).
Furthermore, they are independent of whether we have a bino-- or
wino--like $\neut_1$ scenario\footnote{In the case of a higgsino--like
$\neut_1$, one of the peaks in the lepton-$p_T$ spectrum would be more
pronounced since we get twice as many leptons from the neutralino
decays compared to the bino-- and wino--like $\neut_1$ scenarios.}.
However, note that the absolute number of like--sign dimuon pairs is
different for the scenarios $\textit{S1}$ and $\textit{S2}$, \cf
Sect.~\ref{Sect:decays}.

The signal acceptance $\mathcal{A}$ of the like--sign prompt dimuon search is
evaluated by simulating the process $pp\to\slepton^*/\ssnu^*\to(\mu/\nu) \neut_1$ in {\tt
Herwig6.510}. We simulated 5000 events for each point in the ($m_{\neut
_1},\smass$) mass plane, where we use step sizes of $\Delta\smass=10
\gev$ and $\Delta m_{\neut_1}=20\gev$. For $m_{\neut_1}\le40\gev$ 
(light neutralino) and $m_{\neut_1}\in\{\smass-40\gev,~\smass\}$ 
(boundary region), we decrease the neutralino mass step size to $\Delta
m_{\neut_1} = 5\gev$ since the acceptance is rapidly changing in these
regions. The acceptance maps of the four signal regions ($m_{\mu\mu}>
25\gev,~100\gev,~200\gev,~300\gev$) are given in
Fig.~\ref{Fig:acceptance} in Appendix~\ref{App:acceptance}. For large
parts of the ($m_{\neut_1},\smass$) mass plane the acceptance
$\mathcal{A}$ lies between $2\%$ and $7\%$. However, in the regions
with low neutralino masses, $m_{\neut_1} \lesssim (100-200)\gev$, and
in the region with small $\Delta m = \smass - m_{\neut_1}$, the search
becomes insensitive ($\mathcal{A} \lesssim 2\%$), as expected from the
discussion above. More details are given in
Appendix~\ref{App:acceptance}.

The branching ratios $\mathcal{B}(\slepton \to \ell \neut_1)$ and
$\mathcal{B}(\ssnu \to \nu \neut_1)$ are calculated with {\tt
ISAJET7.64} in the same grid for different values of $\lamp$ for both
the bino- and wino-like $\neut_1$ scenario.

The expected signal rate for a given coupling $\lamp_{2jk}$ and masses
$\smass$, $m_{\neut_1}$ is calculated by
\begin{align}
\big[ \sigma^\mathrm{NLO}(\slepton \to \ell \neut_1) &\times \mathcal{B}
(\slepton \to \ell \neut_1) + \nonumber\\
\sigma^\mathrm{NLO}(\ssnu \to \nu \neut_1) &\times \mathcal{B}(\ssnu
\to \nu \neut_1) \big] \times \mathcal{A}(\smass, m_{\neut_1}),
\label{Eq:signalrate}
\end{align} 
where the branching ratios encode the model dependence (on the bino-
or wino-like $\neut_1$ scenario).  The $95\%$ C.L. upper limits on the
fiducial cross section for like--sign dimuon production provided by
\texttt{ATLAS} are $170.24\fb$, $15.68\fb$, $4.76\fb$ and $2.8\fb$ for
the signal regions $m_{\mu\mu} > 25\gev,~100\gev,~200\gev,~300\gev$,
respectively~\cite{ATLAS-CONF-2011-126}. If the signal rate,
Eq.~\eqref{Eq:signalrate}, exceeds the limit in at least one of the
signal regions, we consider the model as excluded.

We estimate the total uncertainty of the theory prediction to
be $10\%$, taking into account a $5\%$ systematic uncertainty
for the parton density functions, $3\%$ from factorization and
renormalization scale uncertainties of the NLO cross
section~\cite{Dreiner:2006sv} and an averaged statistical uncertainty
of the acceptance estimate\footnote{With 5000 simulated events, the
relative statistical uncertainty on a typical value of the acceptance
$\mathcal{A}=1\%~(7\%)$ is $\Delta \mathcal{A}=14\%~(5\%)$.}. In order
to be conservative, we reduce our signal estimate by the $10\%$
uncertainty in the limit setting procedure.

\begin{figure*}
\centering
\subfigure[Upper limits on $\lamp_{211}$.]{
\scalebox{	0.645}{
\input{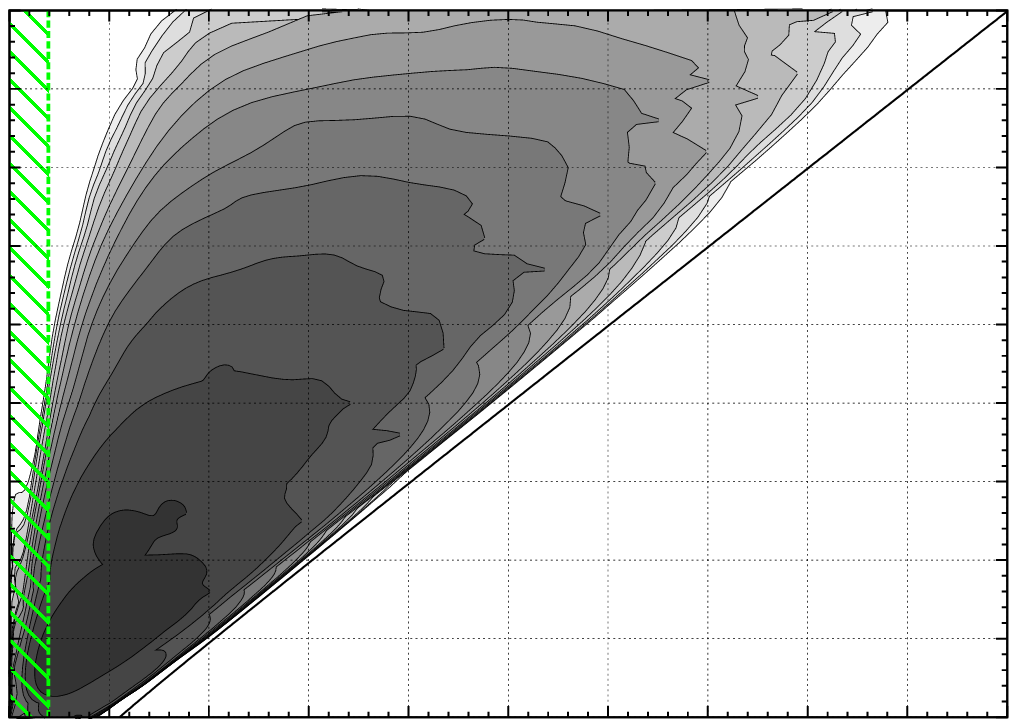}
}}
\subfigure[Upper limits on $\lamp_{212}$.]{
\scalebox{	0.645}{
\input{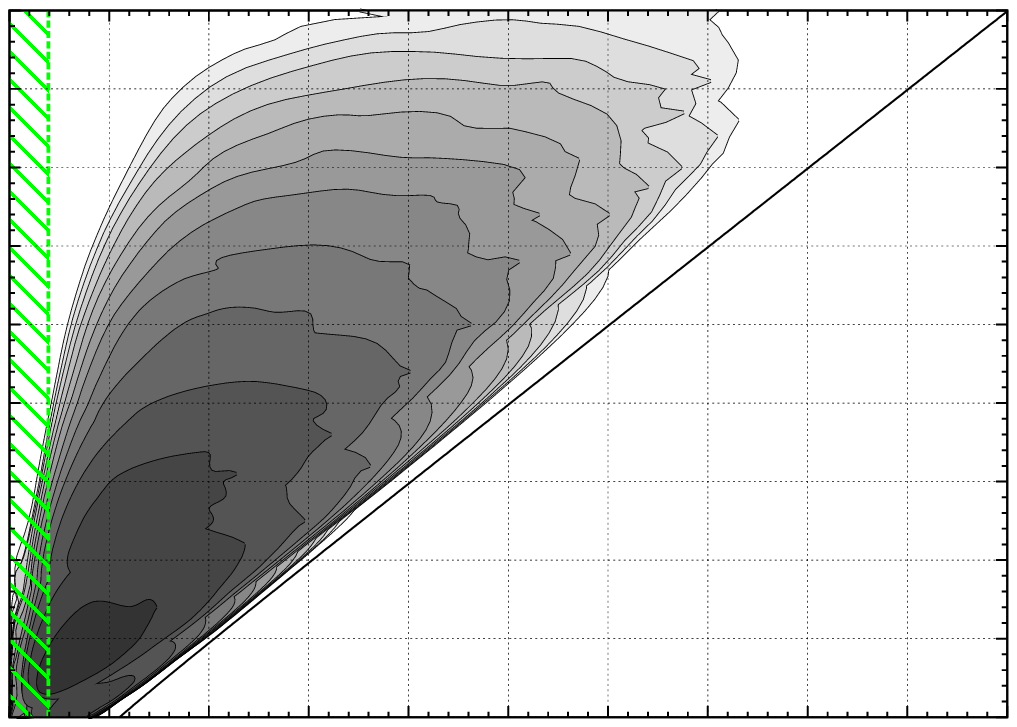}
}}
\subfigure[Upper limits on $\lamp_{221}$.]{
\scalebox{	0.645}{
\input{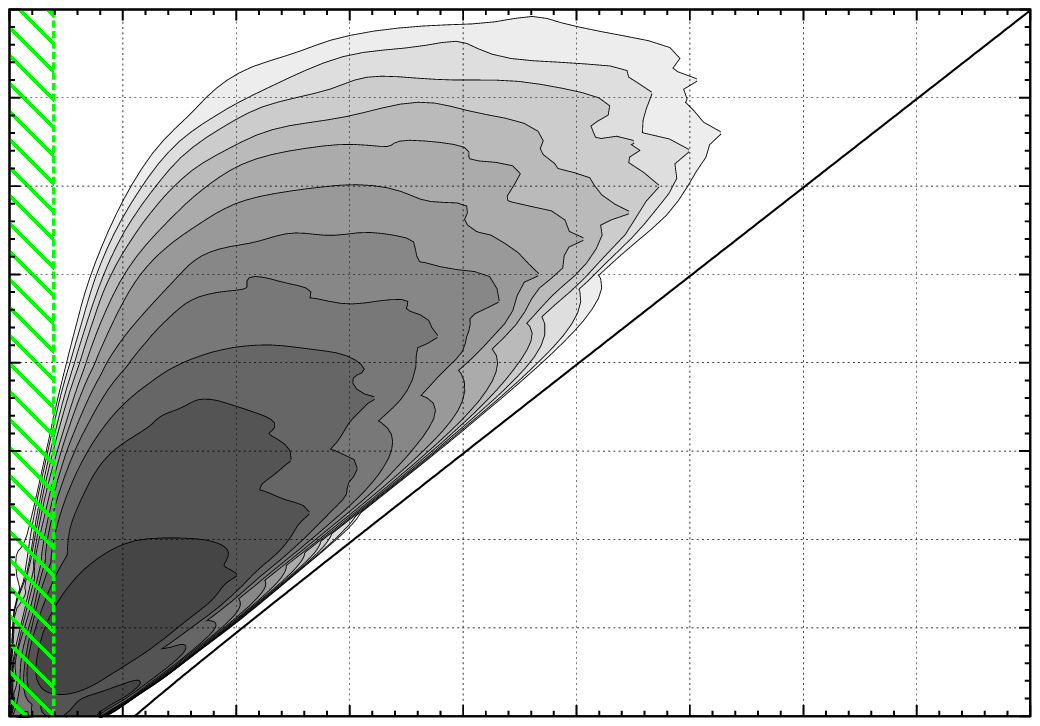}
}}
\subfigure[Upper limits on $\lamp_{222}$.]{
\scalebox{	0.645}{
\input{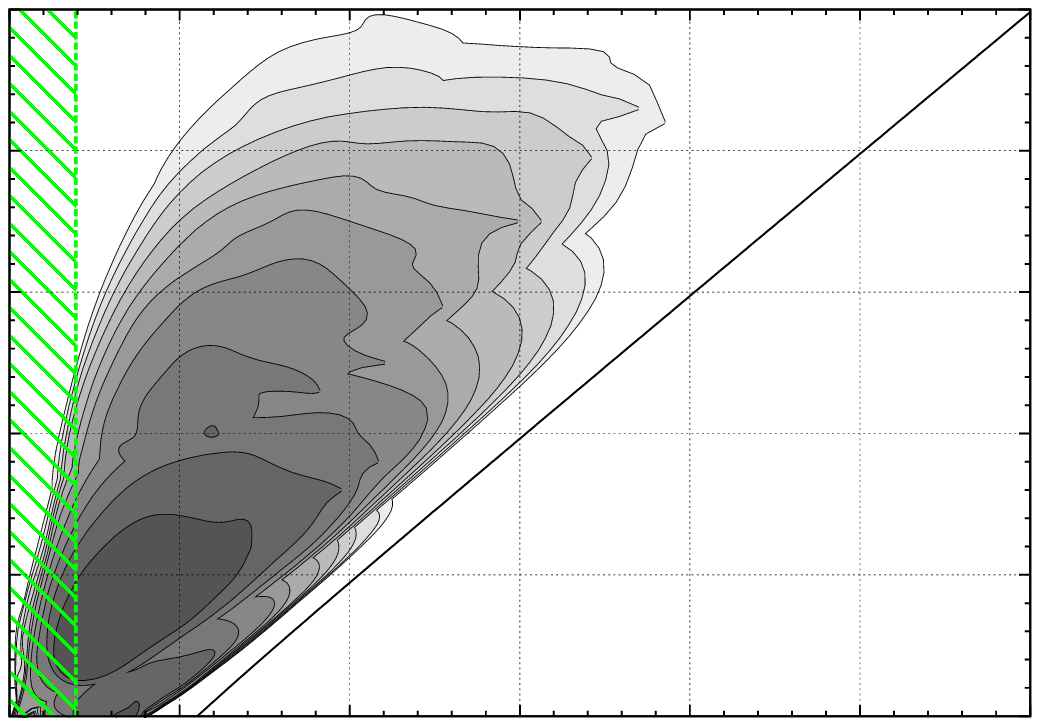}
}}
\caption{Upper bounds on $\lamp_{2jk}~(j,k\in\{1,2\})$ in the $(m_{
\neut_1},\smass)$ mass plane in the bino-like $\neut_1$ scenario,
derived from the \texttt{ATLAS} prompt like--sign dimuon search. The
contour levels are given in steps of $0.0005$. The green striped 
region is excluded due to the lower mass bound from LEP on the lightest
neutralino, $m_{\neut_1} \ge 39\gev$~\cite{Heister:2002jc,Barbier:2004ez}.}
\label{Fig:Limits_bino}
\end{figure*}

\begin{figure*}
\centering
\subfigure[Upper limits on $\lamp_{211}$.]{
\scalebox{	0.645}{
\input{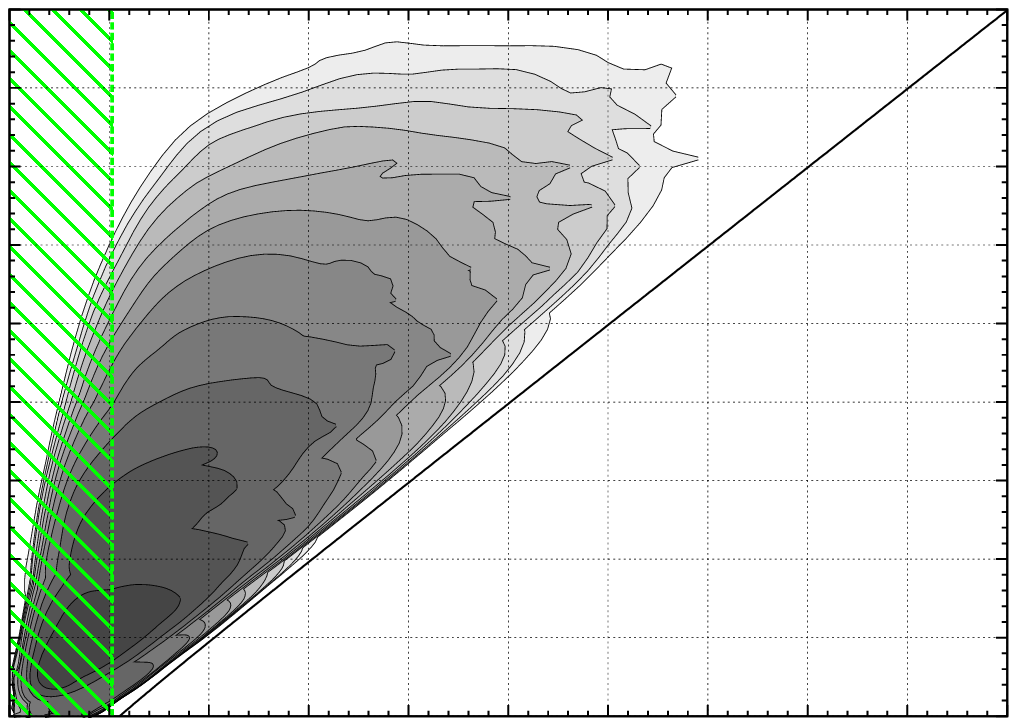}
}}
\subfigure[Upper limits on $\lamp_{212}$.]{
\scalebox{	0.645}{
\input{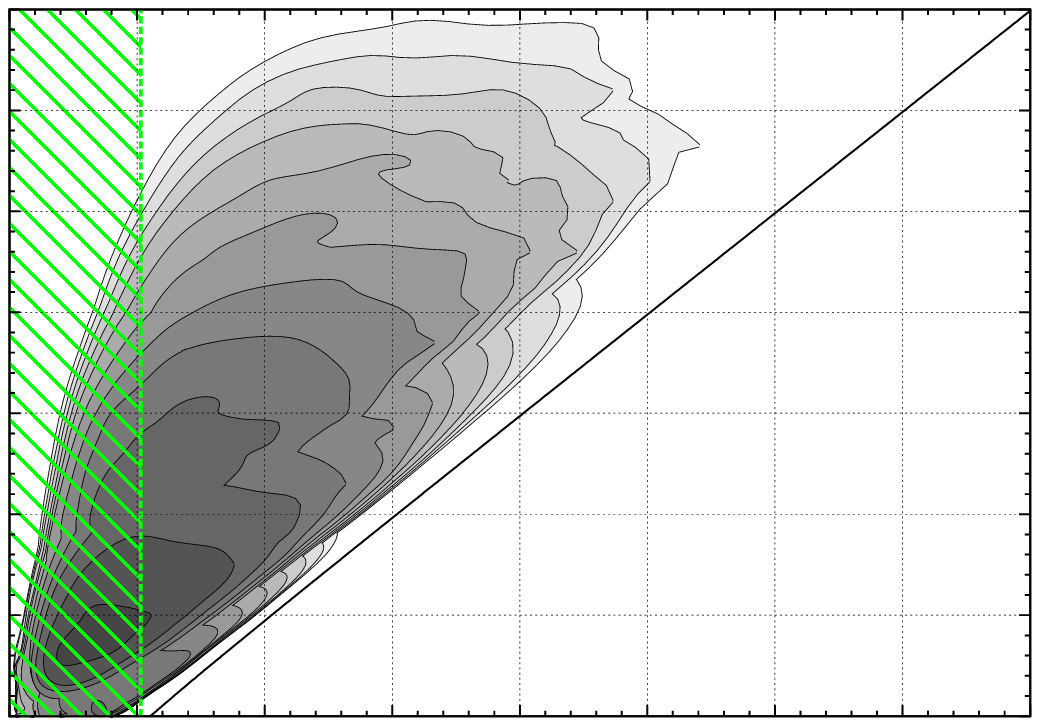}
}}
\subfigure[Upper limits on $\lamp_{221}$.]{
\scalebox{	0.645}{
\input{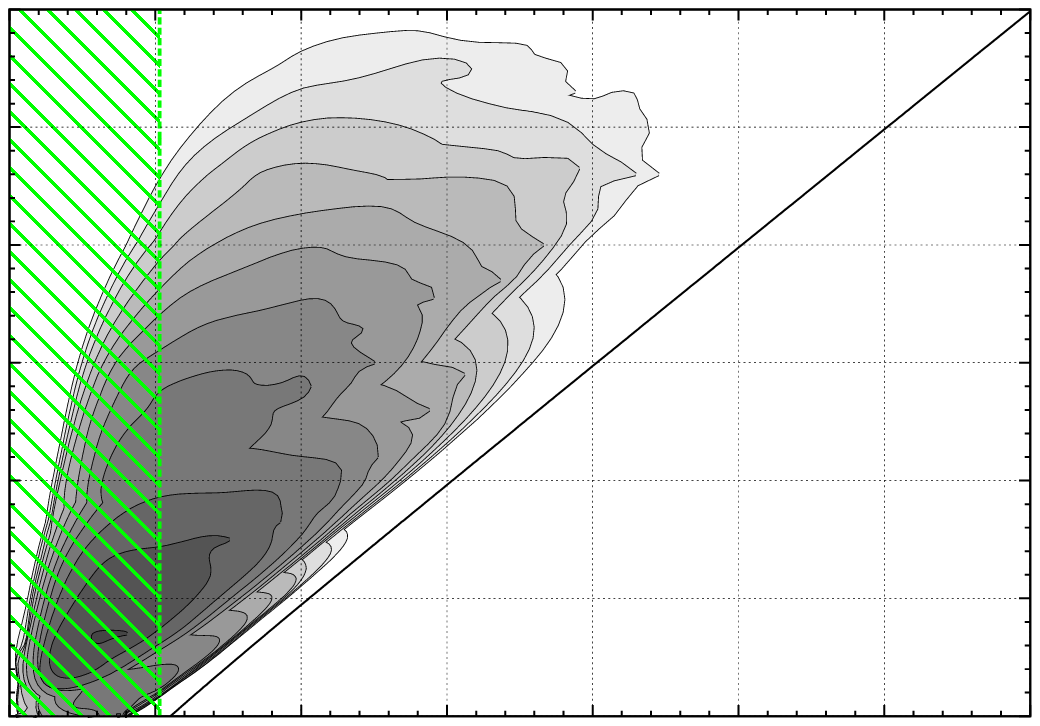}
}}
\subfigure[Upper limits on $\lamp_{222}$.]{
\scalebox{	0.645}{
\input{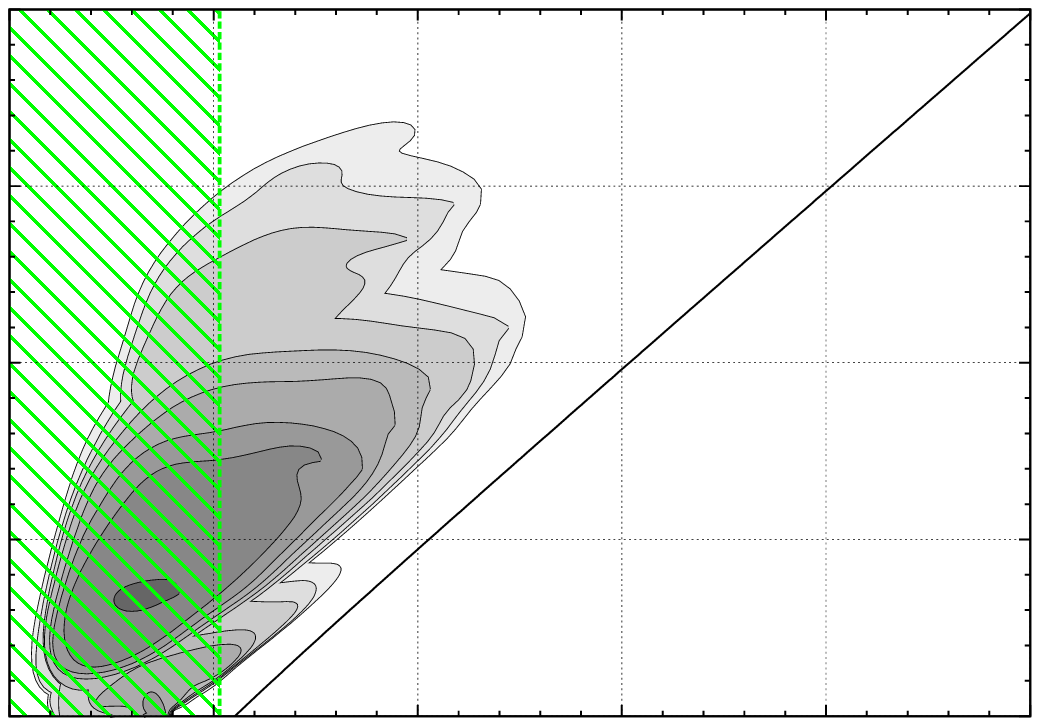}
}}
\caption{Upper bounds on $\lamp_{2jk}~(j,k\in\{1,2\})$ in the 
$(m_{\neut_1},\smass)$ mass plane in the wino-like $\neut_1$ scenario,
derived from the \texttt{ATLAS} prompt like--sign dimuon search. The
contour levels are given in steps of $0.0005$. The green shaded
region is excluded due to the lower mass bound from LEP on the
lightest chargino, $m_{\charge_1^\pm} \ge 103\gev$~\cite{Heister:2002jc,Barbier:2004ez},
which is nearly mass degenerate with the lightest neutralino 
in these scenarios.}
\label{Fig:Limits_wino}
\end{figure*}

We present the upper limits\footnote{Due to our rather simple
treatment of the systematic uncertainties of the signal we cannot
claim our upper limits to be exactly at $95\%$ C.L.. In fact, due to
the conservative approach of subtracting the systematic uncertainty
from the signal yield, we expect our upper limit to be ``at $95\%$
C.L. or more''.} on the four investigated R-parity violating couplings
$\lamp_{2jk}~(j,k\in\{1,2\})$ within the bino--like $\neut_1$
scenario (\textit{S1}) in Fig.~\ref{Fig:Limits_bino}. They are
presented as contours in the ($m_{\neut_1},\smass$) mass plane. The
green striped region indicates the LEP lower mass limit on the
lightest neutralino, $m_{\neut_1}\ge 39\gev$~\cite{Heister:2002jc,Barbier:2004ez}. Note, that this limit (and the limit on the chargino mass) is parameter dependent, \cf Sect.~\ref{Sect:Intro}.

The derived upper bounds on $\lamp$ range from $0.001$ (dark) to
$0.0065$ (bright) and are displayed in steps of $0.0005$ in
grayscale. Since the single slepton production cross section decreases
with the slepton mass, the bounds become weaker for heavier
smuons. Also, due to the insensitivity of the like--sign dimuon search
in the regions of low neutralino mass and low $\Delta m = \smass -
m_{\neut_1}$, we cannot obtain upper bounds on $\lamp$ in these
regions.

The most stringent limits are obtained for the coupling $\lamp_{211}$
due to the larger cross section, \cf Fig.~\ref{Fig:CS}. For a roughly
elliptic region with $m_{\neut_1} \sim \smass -100\gev$ and $\smass
\sim (150 - 300)\gev$, we obtain $\lamp_{211} \le 0.001$. Even for
large smuon masses of $\lesssim\mathcal{O} (1\tev)$, we can still
derive bounds down to $\lamp_{211} \lesssim 0.0045$. The other
couplings are less constrained due to the smaller cross section, \cf
Sect.~\ref{Sect:Production}. The weakest bounds are therefore set on
$\lamp_{222}$, ranging from $0.002$ for $(m_{\neut_1},\smass) \sim
(100, 200)\gev$ to $0.0065$ for smuon masses $\smass \lesssim
550\gev$.

We now turn to the discussion of the results in the wino--like
$\neut_1$ scenario (\textit{S2}) shown in
Fig.~\ref{Fig:Limits_wino}. The LEP lower mass limit on the chargino, $m_{\charge_1^\pm} \ge 103\gev$~\cite{Heister:2002jc,Barbier:2004ez}, is indicated by the green striped region. As discussed in Sect.~\ref{Sect:decays},
we expect only $1/12$ of the time like-sign dimuon events from the
charged slepton gauge decays. Thus, the upper limits on the R-parity
violating coupling $\lamp$ are weaker. For instance, for light smuon
and neutralino masses, $(m_{\neut_1},\smass) = (100, 200)\gev$, the
upper bounds obtained in the wino--like $\neut_1$ scenario are $\lamp
_{211},\lamp_{212} \le 0.0015$, $\lamp_{221} \le 0.002$ and $\lamp_
{222}\le0.0035$.

The bino-like and wino-like $\neut_1$ limits can be interpreted as the
best-case and worst-case scenarios for the like--sign dilepton
signature, respectively. These new limits improve
current limits from the Tevatron~\cite{Abazov:2006ii,Autermann:2006iu}
on $\lamp_{211}$ by a factor $\mathcal{O}(40)$ or more.

We do not consider a higgsino-like lightest neutralino
(\textit{S3}). As discussed in Sect.~\ref{Sect:decays}, the slepton
decay to the higgsino-like $\neut_1$, $\neut_2$ and $\charge_1^\pm$ is
highly suppressed due to the small Yukawa coupling and the competing
R-parity violating decay $\ssmu \to jj$ would dominate,
leading to an overall suppression of the like--sign dimuon signature.
However, we want to remark that exploring the higgsino-like $\neut_1$
scenario with R-parity violating couplings $\lamp_{3jk}$ and a
resonantly produced (left-handed) $\sstau_1$ would be feasible with a
like--sign ditau search.

\section{Conclusions}

We have investigated the impact of LHC data on the resonant production
of single sleptons in R-parity violating models. We presented the NLO
production cross section for resonant sleptons in $pp$ collisions at
$7\tev$ center-of-mass energy. We then discussed the decay modes of
the slepton for three simplified models, where the lightest neutralino
is either bino-- (\textit{S1}), wino--(\textit{S2}), or higgsino--like
(\textit{S3}). We estimated the event yield with a like--sign dilepton
final state. Although these scenarios are simplified, they still
represent wide regions of (realistic) GUT-based SUSY breaking
scenarios like e.g. the CMSSM or the AMSB, as long as the assumed
(relevant) sparticle mass hierarchy is fulfilled.

The main part of this work focused on the derivation of upper bounds
on the R-parity violating couplings from recently published LHC
results. First we considered the dijet signature of resonant sleptons.
Using \texttt{ATLAS} and \texttt{CMS} dijet searches each with
$1\invfb$ of data, we derived upper bounds on the R-parity violating
coupling squared, ${\lamp}_{ijk}^2$ ($i=1,2,3,~j,k=1,2$), times the
branching fraction of the slepton to dijets. These limits only depend
on the mass of the resonant slepton, $\smass$, and are thus
complimentary to low--energy upper bounds, which usually scale
with the squark masses. The limits derived from the \texttt{CMS}
search turn out to be considerably stricter than those of
\texttt{ATLAS}. If the dijet channel is the dominant decay mode,
$\mathcal{B}(\slepton_i/\ssnu_i\to jj)\approx100\%$, the upper bounds
obtained are for instance $\lamp_{i11} \le 0.03~(0.05)$ and
$\lamp_{i22} \le 0.18~(0.37)$ for a slepton mass $\smass =
1000\gev~(1500\gev)$. The complete \texttt{ATLAS} and \texttt{CMS}
results are listed in Tab.~\ref{Tab:ATLASdijet}
and~\ref{Tab:CMSdijet}, respectively. However, these limits from LHC
dijet resonance searches only apply for a very massive spectrum where
the slepton mass is in the range $0.9\tev \le \smass \le 2.5\tev$,
since a dijet resonance search in the lower mass region is still
insensitive due to the overwhelming QCD background.

We then studied the like--sign dilepton signature, which is a very
promising channel for resonant slepton production due to the small SM
background. Using an \texttt{ATLAS} search for anomalous like-sign
dimuon pairs with $1.6\invfb$ of data, we set limits on $\lamp_{211}$,
$\lamp_{212}$, $\lamp_{221}$ and $\lamp_{222}$ in the lightest
neutralino--slepton mass plane, ($m_{\neut_1},\smass$), assuming a
bino--like (\textit{S1}) or wino--like (\textit{S2}) lightest
neutralino LSP.  These bounds range from $0.001$ (for
low slepton and neutralino masses $\sim (100 - 300)\gev$ in
\textit{S1}) to $0.0065$ (heavier slepton and lightest neutralino
masses up to $1\tev$). The strictest bounds are obtained for the
$\lamp_{211}$ coupling for a bino--like lightest neutralino
(\textit{S1}). Our results improve the bounds on $\lamp_{211}$
obtained from the Tevatron by a factor $\gtrsim \mathcal{O}(40)$. For instance, for a slepton mass $\smass = 300~(400)\gev$ and a neutralino mass $m_{\neut_1} = 150~(200)\gev$, the upper bound $\lamp_{211} < 0.04~(0.08)$ obtained by \texttt{D\O}~\cite{Abazov:2006ii,Autermann:2006iu} has improved to $0.001~(0.0015)$ by our analysis of the LHC data.

Furthermore, we discussed in some detail the performance of the
\texttt{ATLAS} like--sign dimuon search on the resonant slepton
signal. For this, we presented the $p_T$ distribution of the isolated
muons and the like-sign dimuon invariant mass distribution for three
different mass configurations [$\smass=500\gev$, $m_{\neut_1}
=(100,250,400)\gev$]. The signal acceptance is reduced for
(\textit{i}) small neutralino masses and (\textit{ii}) for a
low mass difference between the slepton and the lightest neutralino.
In either case one of the muons has a rather low transverse
momentum.

We want to remark that scalar leptoquark searches at
\texttt{ATLAS}~\cite{Aad:2011uv} and
\texttt{CMS}~\cite{Chatrchyan:2011ar} are also sensitive to resonant slepton production. These
analyses searched for two jets associated with either two leptons or
one lepton and missing energy (coming from a neutrino). As discussed
in Sect.~\ref{Sect:decays}, this is also a typical signature of resonant
slepton production. Furthermore, the analyses with one
final state lepton should perform better than the (like--sign)
dilepton search in the parameter region of small mass difference
between the slepton and the lightest neutralino, where the lepton
detection efficiency is low due to reduced phase-space.

We also want to encourage the \texttt{ATLAS} and \texttt{CMS}
collaborations to perform a similar search for like--sign ditau
pairs. This would shed new light on the R-parity violating couplings
$\lamp_{3ij}~(i,j=1,2)$ assuming a resonantly produced $\sstau_1$ with
non-negligible left-handed component.

\begin{acknowledgements}

We thank Sebastian Grab for helpful discussions, reading the
manuscript and for providing the code for the NLO cross section
computation. Furthermore we are grateful to Phillip Bechtle, Till
F. Eifert, Karl Jacobs and Peter Wienemann for their help on
experimental questions. We thank Ingo Ansbach for the permission to
reproduce his poem here.
This work was partially funded by the Helmholtz Alliance ``Physics at the Terascale'' and by the BMBF ``Verbundprojekt HEP-Theorie'' under the contract 0509PDE. TS thanks the Bonn-Cologne Graduate School of Physics and Astronomy for additional financial support.

\end{acknowledgements}
\appendix

\section{Additional tables for the dijet resonance search results}
\label{App:add}

\begin{table*}
\centering
\renewcommand{\arraystretch}{1.1}\addtolength{\tabcolsep}{0.1cm}
\begin{tabular*}{0.68\textwidth}{@{\extracolsep{\fill}}ccccccc}
\hline\hline
								&						&						&	\multicolumn{4}{c}{Upper limits on ${\lamp}_{ijk}^2 \times \mathcal{B}(\slepton_i / \ssnu_i \to j j)$} \\
$\smass$ [GeV]		&	$\mathcal{A}$ (in \%)	&	$\sigma_G/m_G$ (in \%)	&	$i11$	&	$i12$	&	$i21$	&	$i22$	\\
\hline
$  900$ & $  8.1$ & $  8.1$ & $   0.00226$ & $   0.00497$ & $   0.00953$ & $   0.05931$ \\
$  950$ & $  8.0$ & $  7.2$ & $   0.00329$ & $   0.00734$ & $   0.01432$ & $   0.09274$ \\
$ 1000$ & $  7.9$ & $  7.5$ & $   0.00473$ & $   0.01067$ & $   0.02117$ & $   0.14252$ \\
$ 1050$ & $  8.2$ & $  7.3$ & $   0.00542$ & $   0.01234$ & $   0.02490$ & $   0.17413$ \\
$ 1100$ & $  7.9$ & $  6.3$ & $   0.00483$ & $   0.01110$ & $   0.02275$ & $   0.16507$ \\
$ 1150$ & $  8.6$ & $  7.9$ & $   0.00731$ & $   0.01694$ & $   0.03524$ & $   0.26513$ \\
$ 1200$ & $  8.8$ & $  6.8$ & $   0.00619$ & $   0.01442$ & $   0.03045$ & $   0.23733$ \\
$ 1250$ & $  8.8$ & $  6.5$ & $   0.00754$ & $   0.01764$ & $   0.03779$ & $   0.30482$ \\
$ 1300$ & $  8.6$ & $  7.5$ & $   0.01002$ & $   0.02349$ & $   0.05104$ & $   0.42574$ \\
$ 1350$ & $  9.0$ & $  7.0$ & $   0.00873$ & $   0.02051$ & $   0.04516$ & $   0.38927$ \\
$ 1400$ & $  9.0$ & $  7.5$ & $   0.00871$ & $   0.02044$ & $   0.04560$ & $   0.40587$ \\
$ 1450$ & $  9.1$ & $  6.4$ & $   0.00686$ & $   0.01608$ & $   0.03634$ & $   0.33384$ \\
$ 1500$ & $  9.1$ & $  6.5$ & $   0.00815$ & $   0.01904$ & $   0.04358$ & $   0.41282$ \\
$ 1550$ & $  9.3$ & $  6.3$ & $   0.00924$ & $   0.02149$ & $   0.04976$ & $   0.48586$ \\
$ 1600$ & $  9.5$ & $  6.3$ & $   0.01050$ & $   0.02426$ & $   0.05683$ & $   0.57162$ \\
$ 1650$ & $  9.4$ & $  6.4$ & $   0.01303$ & $   0.02987$ & $   0.07077$ & $   0.73286$ \\
$ 1700$ & $  9.8$ & $  6.3$ & $   0.01364$ & $   0.03098$ & $   0.07419$ & $   0.79059$ \\
$ 1750$ & $  9.4$ & $  6.1$ & $   0.01547$ & $   0.03478$ & $   0.08418$ & $   0.92254$ \\
$ 1800$ & $  9.6$ & $  6.4$ & $   0.01769$ & $   0.03929$ & $   0.09605$ &  -  \\
$ 1850$ & $  9.8$ & $  5.9$ & $   0.02210$ & $   0.04840$ & $   0.11951$ &  -  \\
$ 1900$ & $  9.8$ & $  7.0$ & $   0.03023$ & $   0.06522$ & $   0.16255$ &  -  \\
$ 1950$ & $ 10.0$ & $  6.6$ & $   0.04261$ & $   0.09040$ & $   0.22733$ &  -  \\
$ 2000$ & $ 10.0$ & $  6.0$ & $   0.05815$ & $   0.12116$ & $   0.30730$ &  -  \\
$ 2100$ & $ 10.2$ & $  6.5$ & $   0.11257$ & $   0.22523$ & $   0.58045$ &  -  \\
$ 2200$ & $ 10.4$ & $  5.9$ & $   0.21673$ & $   0.41398$ &  -  &  -  \\
$ 2300$ & $ 10.5$ & $  6.3$ & $   0.41049$ & $   0.74428$ &  -  &  -  \\
$ 2400$ & $ 10.6$ & $  5.7$ & $   0.76454$ &  -  &  -  &  -  \\
\hline\hline
\end{tabular*}
\caption{Upper limits on ${\lamp}^2 \times \mathcal{B}(\slepton_i / \ssnu_i \to 
j j)$ derived from the \texttt{ATLAS} search for dijet resonances. The first 
column gives the resonant slepton mass, $\smass$ (in GeV), the second and 
the third column show the acceptance $\mathcal{A}$ (in \%) and the width-to-mass 
ratio, $\sigma_G / m_G$ (in \%), of the gaussian resonance fit, respectively. The 
other columns contain the upper limits on ${\lamp}^2 \times \mathcal{B}(\slepton_i 
/ \ssnu_i \to j j)$, where the indices of $\lamp$ are indicated in the table header 
($i=1,2,3$).}
\label{Tab:ATLASdijet}
\end{table*}

\begin{table*}
\centering
\renewcommand{\arraystretch}{1.1}\addtolength{\tabcolsep}{0.1cm}
\begin{tabular*}{0.6\textwidth}{@{\extracolsep{\fill}}cccccc}
\hline\hline
								&				&	\multicolumn{4}{c}{Upper limits on ${\lamp}_{ijk}^2 \times \mathcal{B}(\slepton_i / \ssnu_i \to j j)$} \\
$\smass$ [GeV]		&	$\mathcal{A}$ (in \%)	&	$i11$	&	$i12$	&	$i21$	&	$i22$	\\
\hline
$ 1000$ & $ 33.8$ & $   0.00102$ & $   0.00229$ & $   0.00455$ & $   0.03064$ \\
$ 1100$ & $ 34.8$ & $   0.00117$ & $   0.00269$ & $   0.00552$ & $   0.04007$ \\
$ 1200$ & $ 35.7$ & $   0.00163$ & $   0.00380$ & $   0.00803$ & $   0.06254$ \\
$ 1300$ & $ 35.7$ & $   0.00201$ & $   0.00472$ & $   0.01026$ & $   0.08555$ \\
$ 1400$ & $ 36.6$ & $   0.00195$ & $   0.00458$ & $   0.01023$ & $   0.09103$ \\
$ 1500$ & $ 36.6$ & $   0.00275$ & $   0.00642$ & $   0.01469$ & $   0.13914$ \\
$ 1600$ & $ 37.3$ & $   0.00413$ & $   0.00954$ & $   0.02235$ & $   0.22478$ \\
$ 1700$ & $ 37.3$ & $   0.00619$ & $   0.01407$ & $   0.03370$ & $   0.35911$ \\
$ 1800$ & $ 38.1$ & $   0.00766$ & $   0.01701$ & $   0.04160$ & $   0.46863$ \\
$ 1900$ & $ 37.6$ & $   0.01441$ & $   0.03108$ & $   0.07747$ & $   0.92097$ \\
$ 2000$ & $ 38.2$ & $   0.02956$ & $   0.06159$ & $   0.15622$ &  -  \\
$ 2100$ & $ 38.6$ & $   0.05246$ & $   0.10497$ & $   0.27053$ &  -  \\
$ 2200$ & $ 38.2$ & $   0.09454$ & $   0.18058$ & $   0.47210$ &  -  \\
$ 2300$ & $ 39.0$ & $   0.18974$ & $   0.34403$ & $   0.91070$ &  -  \\
$ 2400$ & $ 39.1$ & $   0.39971$ & $   0.68404$ &  -  &  -  \\
$ 2500$ & $ 39.1$ & $   0.82990$ &  -  &  -  &  -  \\
\hline\hline
\end{tabular*}
\caption{Upper limits on ${\lamp}^2 \times \mathcal{B}(\slepton_i / 
\ssnu_i \to j j)$ derived from the \texttt{CMS} search for narrow dijet 
resonances. The first column gives the resonant slepton mass,
$\smass$ (in GeV) and the second shows the acceptance
$\mathcal{A}$ (in \%). The other columns contain the upper limits on
${\lamp}^2 \times \mathcal{B}(\slepton_i / \ssnu_i \to j j)$, where
the indices of $\lamp$ are indicated in the table header ($i=1,2,3$).}
\label{Tab:CMSdijet}
\end{table*}

The results of the dijet resonance study in Sect.~\ref{Sect:2j} are
listed in Tab.~\ref{Tab:ATLASdijet} and~\ref{Tab:CMSdijet} for the
\texttt{ATLAS} and \texttt{CMS} analyses, respectively. The upper
bounds on the R-parity violating coupling squared times the branching
ratio of the slepton to dijets, ${\lamp_{ijk}}^2\times
\mathcal{B}(\slepton_i / \ssnu_i \to j j)$, are presented for all
$j,k\in\{1,2\}$ seperately up to the perturbativity bound. We also give the
signal acceptance $\mathcal{A}$ for each slepton mass $\smass$, which
has been evaluated with our MC simulation. For the \texttt{ATLAS}
results, Tab.~\ref{Tab:ATLASdijet}, we also provide the resonance
width to mass ratio, $\sigma_G/m_G$, as derived from a Gaussian fit to
the resonance.

\section{Signal acceptance of the prompt like--sign dimuon search}
\label{App:acceptance}

\begin{figure*}
\centering
\subfigure[$m_{\mu\mu} > 15\gev$ signal region.]{
\scalebox{0.63}{
\input{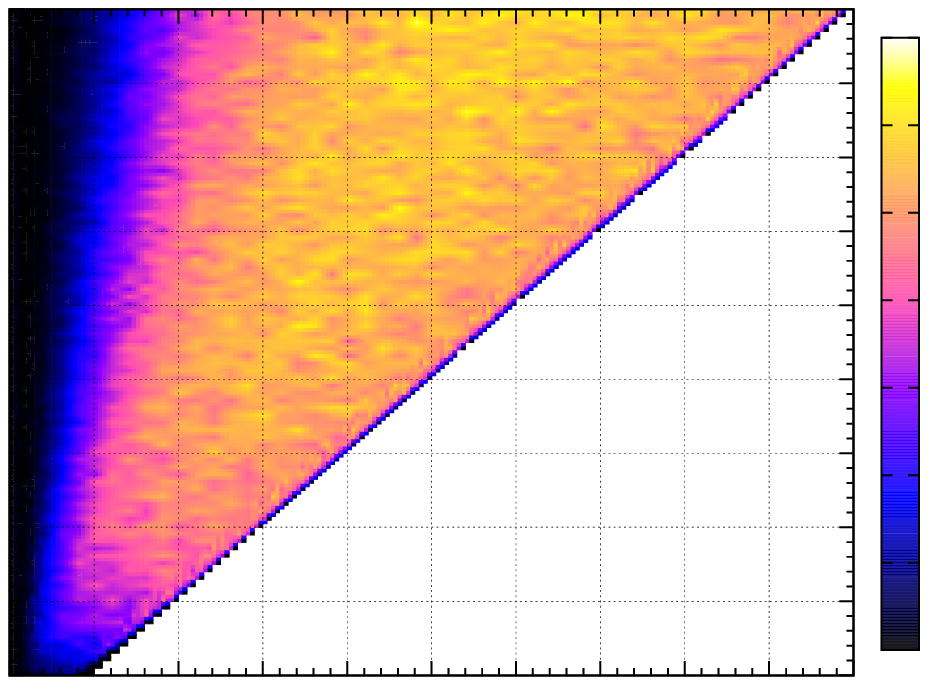}
}}
\hspace{0.2cm}
\subfigure[$m_{\mu\mu} > 100\gev$ signal region.]{
\scalebox{0.63}{
\input{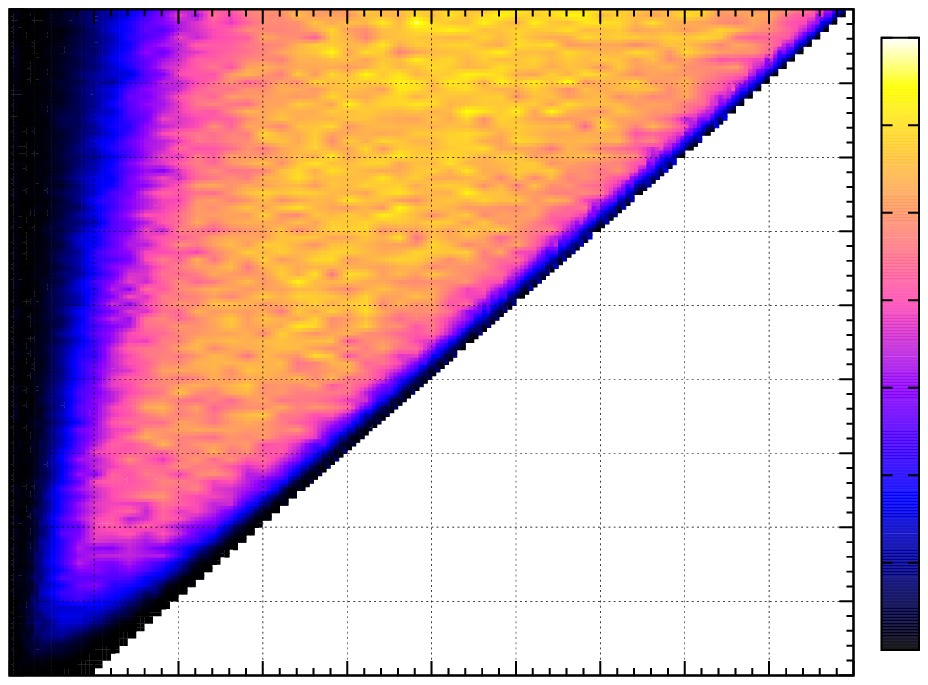}
}}
\subfigure[$m_{\mu\mu} > 200\gev$ signal region.]{
\scalebox{0.63}{
\input{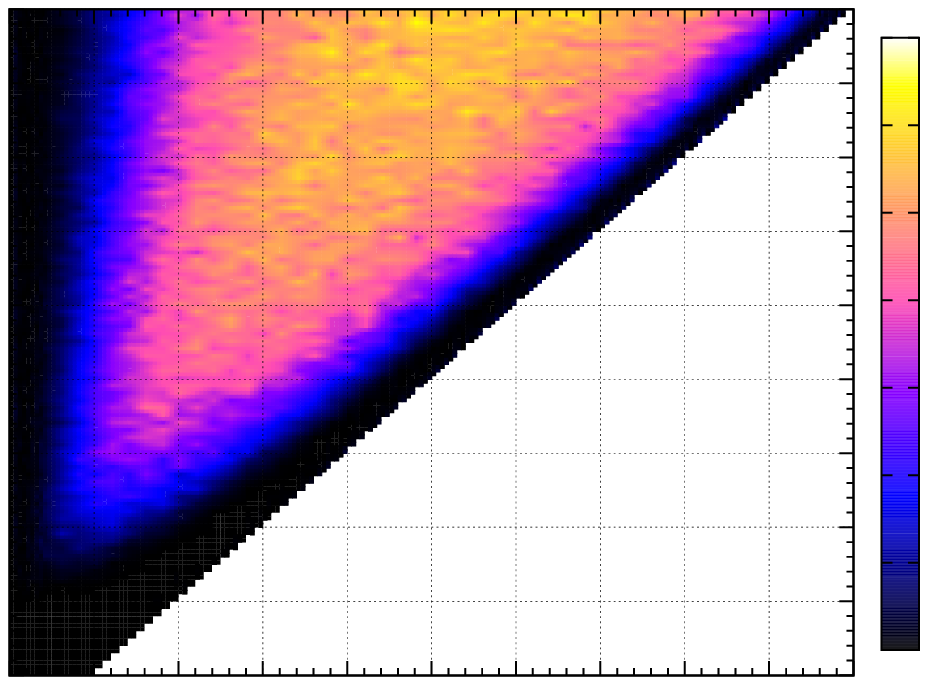}
}}
\hspace{0.2cm}
\subfigure[$m_{\mu\mu} > 300\gev$ signal region.]{
\scalebox{0.63}{
\input{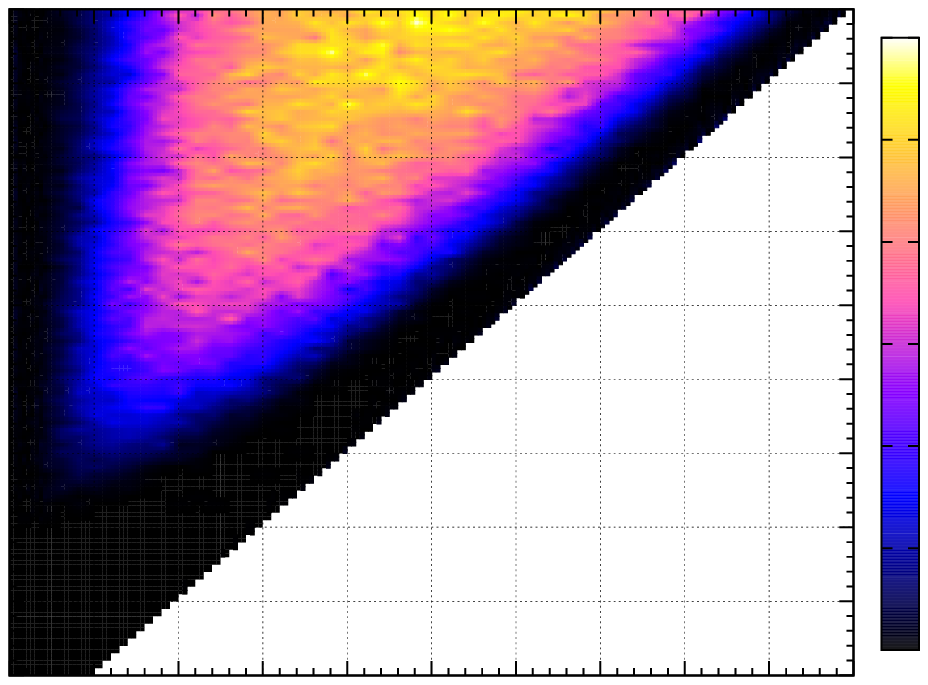}
}}
\caption{Signal acceptance $\mathcal{A}$ of the \texttt{ATLAS} same-sign prompt dimuon search for 
the resonant slepton production process $pp\to \slepton^*/\ssnu^* \to (\mu/\nu)\neut_1$. The subfigures (\textit{a,b,c,d}) show the four signal regions with $m_{\mu\mu} > (25,~100,~200,~300)\gev$, respectively.}
\label{Fig:acceptance}
\end{figure*}

In Fig.~\ref{Fig:acceptance} we give the signal acceptance in the
($m_{\neut_1}, \smass$) mass plane for each signal region ($m_{\mu\mu}
> 25\gev,~100\gev,~200\gev,~300\gev$) of the \texttt{ATLAS} prompt
like--sign dimuon search~\cite{ATLAS-CONF-2011-126} for the simulated
process $pp\to \slepton^*/\ssnu^* \to (\mu/\nu)\neut_1$.

For most of the parameter space, the acceptance ranges between $2\%$
and $7\%$, where the highest largest values are obtained for models
with $m_{\neut_1} \approx \smass/2$. In that case, neither the slepton
nor the neutralino decay are kinematically suppressed, leading to
sizable transverse momenta of the two leptons. In contrast, the
regions with either a low neutralino mass or a low mass difference
between slepton and lightest neutralino, $\Delta m = \smass -
m_{\neut_1}$, feature a very small acceptance. Here, one of the
leptons is soft due to reduced phase space, as discussed in
Sect.~\ref{Sect:likesign2m}, and therefore fails to pass the minimum
$p_T$ requirement.

The insensitive region at low neutralino masses does not depend on the
specific $m_{\mu\mu}$ requirement, since it typically features higher
values of $m_{\mu\mu}$, \cf Fig.~\ref{Fig:Mll}. In contrast, the
acceptance in the low $\Delta m$ region highly depends on the
$m_{\mu\mu}$ cut. Decreasing the mass difference $\Delta m$ leads to a
shift of the $m_{\mu\mu}$ distribution towards lower values. Thus, only the
$m_{\mu\mu}>25\gev$ signal region is capable of exploring the
parameter region with $\Delta m$ down to $\approx 10\gev$, while the
other signal regions with $m_{\mu\mu}>(100,~200,~300)\gev$ require
a mass difference of $\Delta m \gtrsim (20,~75,~150)\gev$,
respectively, to become sensitive (\ie to obtain $\mathcal{A} \gtrsim
2\%$).

Furthermore, in order to obtain a large $m_{\mu\mu}$ value, the
slepton mass $\smass$ has to be sufficiently large. Thus, the signal
regions with $m_{\mu\mu}>(25,~100,~200,~300)\gev$ become sensitive for
slepton masses $\smass \gtrsim (125,~200,~330,~500)\gev$,
respectively.

Although the $m_{\mu\mu} \ge 25\gev$ selection has the best acceptance
coverage, it is still important to use also the other signal regions,
because they have less SM background and thus stricter upper limits on
the fiducial cross section. In parameter regions with heavier sleptons
$\smass\gtrsim\mathcal{O}(600\gev)$ and neutralino masses around
$\smass/2$, the signal region with $m_{\mu\mu} > 300\gev$ typically
poses the strictest limits on the R-parity violating couplings.


\begin{thebibliography}{200}

\bibitem{Gedicht}
During the first run, however there was a problem: \\

Bei Genf trieb man die Teile\\
... mit Energie zur Eile\\
doch wo die Schweizer wohnen\\
das gilt auch f\"ur Protonen\\
dass sie das Hetzen hassen\\
und sich nicht hetzen lassen\\
deshalb dachten die Teilchen\\
wir warten noch ein Weilchen\\
sonst enden wir ja doch\\
in einem schwarzen Loch\\

So blieb das einzig Schnelle\\
am Schluss die offizielle\\
Erkl\"arung der Misere\\
da hie{\ss} es gleich, es w\"are\\
wohl zwischen zwei Magneten\\
ein Schwelbrand aufgetreten\\
die K\"uhlung sei betroffen\\
die Fortsetzung noch offen...\\
als Teilchen alldieweil\\
denkt man sich seinen Teil \\

Ingo Ansbach\\
\textit{Zwielicht zum Preis von einem: Gedichte}\\
Verlag Liber Libri Wien/Pegasus Lyrik, 1. Auflage (2009)

\bibitem{Nilles:1983ge}
For reviews on supersymmetry see for example: 
  H.~P.~Nilles,
  Phys.\ Rept.\  {\bf 110 } (1984)  1-162; 
  H.~E.~Haber and G.~L.~Kane,
  Phys.\ Rept.\  {\bf 117} (1985) 75;
  M.~Drees,
  [hep-ph/9611409];
  S.~P.~Martin,
  In *Kane, G.L. (ed.): Perspectives on supersymmetry II* 1-153
  [hep-ph/9709356].

\bibitem{Farrar:1978xj}
  G.~R.~Farrar, P.~Fayet,
  Phys.\ Lett.\  {\bf B76 } (1978)  575-579.

\bibitem{ATLAS_SUSYMETsearches}
  G.~Aad {\it et al.}  [\texttt{ATLAS} Collaboration],
  arXiv:1109.6572 [hep-ex];
  \idem arXiv:1109.6606 [hep-ex];
  \idem JHEP {\bf 1111} (2011) 099
  [arXiv:1110.2299 [hep-ex]];
  \idem arXiv:1110.6189 [hep-ex];
  \idem arXiv:1111.4116 [hep-ex];
  \idem arXiv:1112.3832 [hep-ex].    

\bibitem{CMS_SUSYMETsearches}
  S.~Chatrchyan {\it et al.}  [\texttt{CMS} Collaboration],
  Phys.\ Rev.\ Lett.\  {\bf 106} (2011) 211802
  [arXiv:1103.0953 [hep-ex]];
  \idem JHEP {\bf 1106} (2011) 026
  [arXiv:1103.1348 [hep-ex]];
  \idem JHEP {\bf 1106} (2011) 077
  [arXiv:1104.3168 [hep-ex]];
  \idem JHEP {\bf 1106} (2011) 093
  [arXiv:1105.3152 [hep-ex]];
  \idem JHEP {\bf 1107} (2011) 113
  [arXiv:1106.3272 [hep-ex]];
  \idem JHEP {\bf 1108} (2011) 155
  [arXiv:1106.4503 [hep-ex]];
  \idem arXiv:1107.1279 [hep-ex];
  \idem JHEP {\bf 1108} (2011) 156
  [arXiv:1107.1870 [hep-ex]];
  \idem arXiv:1109.2352 [hep-ex].

\bibitem{Bechtle:2011dm}
  P.~Bechtle, B.~Sarrazin, K.~Desch, H.~K.~Dreiner, P.~Wienemann, M.~Kramer, C.~Robens, B.~O'Leary,
  Phys.\ Rev.\  {\bf D84 } (2011)  011701.
  [arXiv:1102.4693 [hep-ph]]; 
  O.~Buchmueller, R.~Cavanaugh, A.~De Roeck, M.~J.~Dolan, J.~R.~Ellis, H.~Flacher, S.~Heinemeyer, G.~Isidori {\it et al.},
   [arXiv:1110.3568 [hep-ph]];
   B.~C.~Allanach, T.~J.~Khoo, C.~G.~Lester, S.~L.~Williams,
  JHEP {\bf 1106 } (2011)  035.
  [arXiv:1103.0969 [hep-ph]]; 
  S.~Sekmen, S.~Kraml, J.~Lykken, F.~Moortgat, S.~Padhi, L.~Pape, M.~Pierini and H.~B.~Prosper {\it et al.},
  arXiv:1109.5119 [hep-ph];
  A.~Arbey, M.~Battaglia and F.~Mahmoudi,
  Eur.\ Phys.\ J.\ C {\bf 72} (2012) 1847
  [arXiv:1110.3726 [hep-ph]].

\bibitem{Dreiner:1997uz}
  H.~K.~Dreiner,
  In *Kane, G.L. (ed.): Perspectives on supersymmetry* 462-479.
  [hep-ph/9707435].

\bibitem{Allanach:2003eb}
  B.~C.~Allanach, A.~Dedes, H.~K.~Dreiner,
  Phys.\ Rev.\  {\bf D69 } (2004)  115002.
  [hep-ph/0309196];
  B.~C.~Allanach, A.~Dedes and H.~K.~Dreiner,
  Phys.\ Rev.\ D {\bf 60} (1999) 056002
  [hep-ph/9902251].

\bibitem{Dreiner:2005rd}
  H.~K.~Dreiner, C.~Luhn, M.~Thormeier,
  Phys.\ Rev.\  {\bf D73 } (2006)  075007.
  [hep-ph/0512163]; L.~E.~Ibanez, G.~G.~Ross,
  Phys.\ Lett.\  {\bf B260 } (1991)  291-295.

\bibitem{Hirsch:2000ef}
  M.~Hirsch, M.~A.~Diaz, W.~Porod, J.~C.~Romao, J.~W.~F.~Valle,
  Phys.\ Rev.\  {\bf D62 } (2000)  113008.
  [hep-ph/0004115].

\bibitem{Dreiner:2003hw}
  H.~K.~Dreiner, M.~Thormeier,
  Phys.\ Rev.\  {\bf D69 } (2004)  053002.
  [hep-ph/0305270].

\bibitem{Hall:1983id}
  L.~J.~Hall, M.~Suzuki,
  Nucl.\ Phys.\  {\bf B231 } (1984)  419.

\bibitem{Davidson:2000uc}
  S.~Davidson, M.~Losada,
  JHEP {\bf 0005 } (2000)  021.
  [hep-ph/0005080].

\bibitem{Dreiner:2010ye}
  H.~K.~Dreiner, M.~Hanussek, S.~Grab,
  Phys.\ Rev.\  {\bf D82}, 055027 (2010).
  [arXiv:1005.3309 [hep-ph]].

\bibitem{Minkowski:1977sc}
  P.~Minkowski,
  Phys.\ Lett.\  {\bf B67}, 421 (1977).

\bibitem{Mohapatra:1979ia}
  R.~N.~Mohapatra, G.~Senjanovic,
  Phys.\ Rev.\ Lett.\  {\bf 44 } (1980)  912.

\bibitem{Dreiner:2006xw}
  H.~K.~Dreiner, C.~Luhn, H.~Murayama, M.~Thormeier,
  Nucl.\ Phys.\  {\bf B774 } (2007)  127-167.
  [hep-ph/0610026].

\bibitem{Lee:2010vj}
  H.~-S.~Lee,
  Phys.\ Lett.\  {\bf B704 } (2011)  316-321.
  [arXiv:1007.1040 [hep-ph]].

\bibitem{Dreiner:2011ft}
  H.~K.~Dreiner, M.~Hanussek, J.~-S.~Kim, C.~H.~Kom,
 [arXiv:1106.4338 [hep-ph]].

\bibitem{Dimopoulos:1988fr}
  S.~Dimopoulos, R.~Esmailzadeh, L.~J.~Hall, G.~D.~Starkman,
  Phys.\ Rev.\  {\bf D41 } (1990)  2099.

\bibitem{Aad:2011fq}
  G.~Aad {\it et al.} [\texttt{ATLAS} Collaboration],
    [arXiv: 1108.6311 [hep-ex]].

\bibitem{Chatrchyan:2011ns}
  S.~Chatrchyan {\it et al.}  [\texttt{CMS} Collaboration],
  Phys.\ Lett.\ B {\bf 704} (2011) 123
  [arXiv:1107.4771 [hep-ex]].
  
\bibitem{ATLAS-CONF-2011-126}
 \texttt{ATLAS} collaboration,
 ATLAS-CONF-2011-126,
  \url{http://cdsweb.cern.ch/record/1383790}.
  Recently, \texttt{ATLAS} published this analysis with a slightly modified event selection in~\cite{Aad:2012cg}.

\bibitem{Aad:2012cg}
  G.~Aad {\it et al.}  [\texttt{ATLAS} Collaboration],
  arXiv:1201.1091 [hep-ex].

\bibitem{Hewett:1998fu}
  J.~L.~Hewett, T.~G.~Rizzo,
  [hep-ph/9809525].
  
\bibitem{Dreiner:1998gz}
  H.~K.~Dreiner, P.~Richardson, M.~H.~Seymour,
  [hep-ph/9903419].  
  
\bibitem{Dreiner:2000vf}
  H.~K.~Dreiner, P.~Richardson, M.~H.~Seymour,
  Phys.\ Rev.\  {\bf D63 } (2001)  055008.
  [hep-ph/0007228].

\bibitem{Dreiner:2000qf}
  H.~K.~Dreiner, P.~Richardson, M.~H.~Seymour,
  [hep-ph/0001224].

\bibitem{Richardson:2000nt}
  P.~Richardson,
  [hep-ph/0101105].

\bibitem{Deliot:2000mf}
  F.~Deliot, G.~Moreau, C.~Royon,
  Eur.\ Phys.\ J.\  {\bf C19 } (2001)  155-181.
  [hep-ph/0007288].

\bibitem{Dreiner:2006sv}
  H.~K.~Dreiner, S.~Grab, M.~Kramer, M.~K.~Trenkel,
  Phys.\ Rev.\  {\bf D75 } (2007)  035003.
  [hep-ph/0611195].  

\bibitem{Moreau:1999bt}
  G.~Moreau, M.~Chemtob, F.~Deliot, C.~Royon, E.~Perez,
  Phys.\ Lett.\  {\bf B475 } (2000)  184-189.
  [hep-ph/9910341].

\bibitem{Moreau:2000bs}
  G.~Moreau, E.~Perez, G.~Polesello,
  Nucl.\ Phys.\  {\bf B604 } (2001)  3-31.
  [hep-ph/0003012].

\bibitem{Oakes:1997zg}
  R.~J.~Oakes, K.~Whisnant, J.~M.~Yang, B.~-L.~Young, X.~Zhang,
  Phys.\ Rev.\  {\bf D57 } (1998)  534-540.
  [hep-ph/9707477].
  
\bibitem{Bernhardt:2008mz}
  M.~A.~Bernhardt, H.~K.~Dreiner, S.~Grab, P.~Richardson,
  Phys.\ Rev.\  {\bf D78 } (2008)  015016.
  [arXiv:0802.1482 [hep-ph]].

\bibitem{Cakir:2011dx}
  O.~Cakir, S.~Kuday, I.~T.~Cakir, S.~Sultansoy,
  [arXiv:1103.5087 [hep-ph]].  

\bibitem{Allanach:2003wz}
  B.~C.~Allanach, M.~Guchait, K.~Sridhar,
  Phys.\ Lett.\  {\bf B586 } (2004)  373-381.
  [hep-ph/0311254].

\bibitem{Dreiner:2008rv}
  H.~K.~Dreiner, S.~Grab, M.~K.~Trenkel,
  Phys.\ Rev.\  {\bf D79 } (2009)  016002.
  [arXiv:0808.3079 [hep-ph]].  
  
\bibitem{Kilic:2011sr}
  C.~Kilic and S.~Thomas,
  Phys.\ Rev.\ D {\bf 84} (2011) 055012
  [arXiv:1104.1002 [hep-ph]].

\bibitem{Abazov:2006ii}
  V.~M.~Abazov {\it et al.} [\texttt{D\O} Collaboration],
  Phys.\ Rev.\ Lett.\  {\bf 97 } (2006)  111801.
  [hep-ex/0605010].

\bibitem{Autermann:2006iu}
  C.~T.~Autermann,
  FERMILAB-THESIS-2006-46.

\bibitem{Abazov:2007zz}
  V.~M.~Abazov {\it et al.} [\texttt{D\O} Collaboration],
  Phys.\ Rev.\ Lett.\  {\bf 100 } (2008)  241803.
  [arXiv:0711.3207 [hep-ex]]; 
  \idem
  Phys.\ Rev.\ Lett.\ \ {\bf 105} (2010) 191802
  [arXiv:1007.4835 [hep-ex]].

\bibitem{Abulencia:2006xm}
  A.~Abulencia {\it et al.}  [\texttt{CDF} Collaboration],
  Phys.\ Rev.\ Lett.\  {\bf 96} (2006) 211802
  [hep-ex/0603006].

\bibitem{Aaltonen:2010fv}
  T.~Aaltonen {\it et al.}  [\texttt{CDF} Collaboration],
  Phys.\ Rev.\ Lett.\  {\bf 105} (2010) 191801
  [arXiv:1004.3042 [hep-ex]].

\bibitem{Abulencia:2005nf}
  A.~Abulencia {\it et al.}  [\texttt{CDF} Collaboration],
  Phys.\ Rev.\ Lett.\  {\bf 95} (2005) 252001
  [hep-ex/0507104].

\bibitem{Acosta:2005ij}
  D.~Acosta {\it et al.}  [\texttt{CDF} Collaboration],
  Phys.\ Rev.\ Lett.\  {\bf 95} (2005) 131801
  [hep-ex/0506034].
  
\bibitem{Chamseddine:1982jx}
  A.~H.~Chamseddine, R.~L.~Arnowitt and P.~Nath,
  Phys.\ Rev.\ Lett.\  {\bf 49} (1982) 970.

\bibitem{Barbieri:1982eh}
  R.~Barbieri, S.~Ferrara and C.~A.~Savoy,
  Phys.\ Lett.\ B {\bf 119} (1982) 343.
  
\bibitem{Hall:1983iz}
  L.~J.~Hall, J.~D.~Lykken and S.~Weinberg,
  Phys.\ Rev.\ D {\bf 27} (1983) 2359.
  
\bibitem{Kane:1993td}
  G.~L.~Kane, C.~F.~Kolda, L.~Roszkowski and J.~D.~Wells,
  Phys.\ Rev.\ D {\bf 49} (1994) 6173
  [hep-ph/9312272].
  
\bibitem{Abulencia:2007mp}
  A.~Abulencia {\it et al.}  [\texttt{CDF} Collaboration],
  Phys.\ Rev.\ Lett.\  {\bf 98} (2007) 131804
  [arXiv:0706.4448 [hep-ex]].

\bibitem{Abazov:2006nw}
  V.~M.~Abazov {\it et al.} [\texttt{D\O} Collaboration],
  Phys.\ Lett.\  {\bf B638 } (2006)  441-449.
  [hep-ex/0605005].

\bibitem{Brigliadori:2008vf}
  T.~Aaltonen {\it et al.}  [\texttt{CDF} Collaboration],
  Phys.\ Rev.\ Lett.\  {\bf 101} (2008) 071802
  [arXiv:0802.3887 [hep-ex]].

\bibitem{Chakrabarti:2004yq}
  S.~Chakrabarti, M.~Guchait and N.~K.~Mondal,
  Phys.\ Lett.\ B {\bf 600} (2004) 231
  [hep-ph/0404261].

\bibitem{Das:2005mr}
  S.~P.~Das, A.~Datta and S.~Poddar,
  Phys.\ Rev.\ D {\bf 73} (2006) 075014
  [hep-ph/0509171].

\bibitem{Dreiner:2011xa}
  H.~K.~Dreiner, S.~Grab, T.~Stefaniak,
  Phys.\ Rev.\  {\bf D84 } (2011)  015005.
  [arXiv:1103.1883 [hep-ph]].

\bibitem{Butterworth:1992tc}
  J.~Butterworth and H.~K.~Dreiner,
  Nucl.\ Phys.\ B {\bf 397} (1993) 3
  [hep-ph/9211204];
H.~K.~Dreiner and P.~Morawitz,
  Nucl.\ Phys.\ B {\bf 503} (1997) 55
  [hep-ph/9703279].


\bibitem{Aktas:2004ij}
T.~Ahmed {\it et al.}  [\texttt{H1} Collaboration],
  Z.\ Phys.\ C {\bf 64} (1994) 545;
  S.~Aid {\it et al.}  [\texttt{H1} Collaboration],
  Z.\ Phys.\ C {\bf 71} (1996) 211
  [hep-ex/9604006].
  A.~Aktas {\it et al.}  [\texttt{H1} Collaboration],
  Eur.\ Phys.\ J.\ C {\bf 36} (2004) 425
  [hep-ex/0403027].

\bibitem{zeus:2006je}
J.~Breitweg {\it et al.}  [\texttt{ZEUS} Collaboration],
  Eur.\ Phys.\ J.\ C {\bf 16} (2000) 253
  [hep-ex/0002038];
S.~Chekanov {\it et al.}  [\texttt{ZEUS} Collaboration],
  Phys.\ Rev.\ D {\bf 68} (2003) 052004
  [hep-ex/0304008];
  S.~Chekanov {\it et al.}  [\texttt{ZEUS} Collaboration],
  Eur.\ Phys.\ J.\ C {\bf 50} (2007) 269
  [hep-ex/0611018].

\bibitem{Collaboration:2011qr}
  \texttt{ATLAS} Collaboration,
  Eur.\ Phys.\ J.\ C {\bf 71} (2011) 1809
  [arXiv:1109.3089 [hep-ex]].

\bibitem{Aad:2011zb}
  G.~Aad {\it et al.}  [\texttt{ATLAS} Collaboration],
  Phys.\ Lett.\ B {\bf 707} (2012) 478
  [arXiv:1109.2242 [hep-ex]].
  
\bibitem{Chatrchyan:2011ff}
  S.~Chatrchyan {\it et al.}  [\texttt{CMS} Collaboration],
  Phys.\ Lett.\ B {\bf 704} (2011) 411
  [arXiv:1106.0933 [hep-ex]].
  
\bibitem{CMS-PAS-EXO-11-045}
  S.~Chatrchyan {\it et al.}  [\texttt{CMS} Collaboration],
  CMS PAS EXO-11-045,
  \url{http://cdsweb.cern.ch/record/1393758/}.

\bibitem{ATLAS:2011ad}
  G.~Aad {\it et al.}  [\texttt{ATLAS} Collaboration],
  Phys.\ Rev.\ D {\bf 85} (2012) 012006
  [arXiv:1109.6606 [hep-ex]].
  
\bibitem{Banks:1995by}
  T.~Banks, Y.~Grossman, E.~Nardi, Y.~Nir,
  Phys.\ Rev.\  {\bf D52 } (1995)  5319-5325.
  [hep-ph/9505248].

\bibitem{Heister:2002jc}
  A.~Heister {\it et al.}  [\texttt{ALEPH} Collaboration],
  Eur.\ Phys.\ J.\ C {\bf 31} (2003) 1
  [hep-ex/0210014].
  
\bibitem{Barbier:2004ez}
  R.~Barbier, C.~Berat, M.~Besancon, M.~Chemtob, A.~Deandrea, E.~Dudas, P.~Fayet and S.~Lavignac {\it et al.},
  Phys.\ Rept.\  {\bf 420} (2005) 1
  [hep-ph/0406039].

\bibitem{Bhattacharyya:1996nj}
  G.~Bhattacharyya,
  Nucl.\ Phys.\ Proc.\ Suppl.\  {\bf 52A} (1997) 83
  [hep-ph/9608415].

\bibitem{Allanach:1999ic}
  B.~C.~Allanach, A.~Dedes and H.~K.~Dreiner,
  Phys.\ Rev.\ D {\bf 60} (1999) 075014
  [hep-ph/9906209].

\bibitem{Kao:2009fg}
  Y.~Kao and T.~Takeuchi,
  arXiv:0910.4980 [hep-ph].

\bibitem{Bernhardt:2008jz}
  M.~A.~Bernhardt, S.~P.~Das, H.~K.~Dreiner and S.~Grab,
  Phys.\ Rev.\  D {\bf 79} (2009) 035003
  [arXiv:0810.3423 [hep-ph]].

\bibitem{Baer:1989hr}
  H.~Baer, X.~Tata, J.~Woodside,
  Phys.\ Rev.\  {\bf D41 } (1990)  906.

\bibitem{Barnett:1993ea}
  R.~M.~Barnett, J.~F.~Gunion, H.~E.~Haber,
  Phys.\ Lett.\  {\bf B315 } (1993)  349-354.
  [hep-ph/9306204].
    
\bibitem{Dreiner:1993ba}
  H.~K.~Dreiner, M.~Guchait, D.~P.~Roy,
  Phys.\ Rev.\  {\bf D49 } (1994)  3270-3282.
  [hep-ph/9310291].  

\bibitem{Choudhury:2002aua}
  D.~Choudhury, S.~Majhi, V.~Ravindran,
  Nucl.\ Phys.\  {\bf B660 } (2003)  343-361.
  [hep-ph/0207247].

\bibitem{Yang:2005ts}
  L.~L.~Yang, C.~S.~Li, J.~J.~Liu, Q.~Li,
  Phys.\ Rev.\  {\bf D72 } (2005)  074026.
  [hep-ph/0507331].

\bibitem{Chen:2006ep}
  Y.~-Q.~Chen, T.~Han, Z.~-G.~Si,
  JHEP {\bf 0705 } (2007)  068.
  [hep-ph/0612076].
      
\bibitem{Pumplin:2005rh}
  J.~Pumplin, A.~Belyaev, J.~Huston, D.~Stump, W.~K.~Tung,
  JHEP {\bf 0602 } (2006)  032.
  [hep-ph/0512167].

\bibitem{Martin:2002dr}
  A.~D.~Martin, R.~G.~Roberts, W.~J.~Stirling, R.~S.~Thorne,
  Phys.\ Lett.\  {\bf B531 } (2002)  216-224.
  [hep-ph/0201127].
  
\bibitem{Martin:2004ir}
  A.~D.~Martin, R.~G.~Roberts, W.~J.~Stirling, R.~S.~Thorne,
  Phys.\ Lett.\  {\bf B604 } (2004)  61-68.
  [hep-ph/0410230].
    
\bibitem{Majhi:2010zg}
  S.~Majhi, P.~Mathews, V.~Ravindran,
  Nucl.\ Phys.\  {\bf B850 } (2011)  287-320.
  [arXiv:1011.6027 [hep-ph]].

\bibitem{Randall:1998uk}
  L.~Randall and R.~Sundrum,
  Nucl.\ Phys.\ B {\bf 557} (1999) 79
  [hep-th/9810155].

\bibitem{Giudice:1998xp}
  G.~F.~Giudice, M.~A.~Luty, H.~Murayama and R.~Rattazzi,
  JHEP {\bf 9812} (1998) 027
  [hep-ph/9810442].

\bibitem{Bagger:1999rd}
  J.~A.~Bagger, T.~Moroi and E.~Poppitz,
  JHEP {\bf 0004} (2000) 009
  [hep-th/9911029].

\bibitem{Baer:2010uy}
  H.~Baer, S.~de Alwis, K.~Givens, S.~Rajagopalan and H.~Summy,
  JHEP {\bf 1005} (2010) 069
  [arXiv:1002.4633 [hep-ph]]. 

\bibitem{Choi:2007ka}
  K.~Choi and H.~P.~Nilles,
  JHEP {\bf 0704} (2007) 006
  [hep-ph/0702146 [HEP-PH]].
  
\bibitem{Haber:1984rc}
  H.~E.~Haber, G.~L.~Kane,
  Phys.\ Rept.\  {\bf 117 } (1985)  75-263.

\bibitem{Dreiner:2008tw}
  H.~K.~Dreiner, H.~E.~Haber, S.~P.~Martin,
  Phys.\ Rept.\  {\bf 494 } (2010)  1-196.
  [arXiv:0812.1594 [hep-ph]].    

\bibitem{Dreiner:1991pe}
  H.~K.~Dreiner, G.~G.~Ross,
  Nucl.\ Phys.\  {\bf B365 } (1991)  597-613.

\bibitem{Paige:2003mg}
  F.~E.~Paige, S.~D.~Protopopescu, H.~Baer and X.~Tata,
  arXiv:hep-ph/0312045.

\bibitem{Dreiner:2008ca}
  H.~K.~Dreiner, S.~Grab,
  Phys.\ Lett.\  {\bf B679 } (2009)  45-50.
  [arXiv:0811.0200 [hep-ph]].

\bibitem{Desch:2010gi}
  K.~Desch, S.~Fleischmann, P.~Wienemann, H.~K.~Dreiner, S.~Grab,
  Phys.\ Rev.\  {\bf D83 } (2011)  015013.
  [arXiv:1008.1580 [hep-ph]].  
  
\bibitem{ISAWIG}
  \url{http://www.hep.phy.cam.ac.uk/~richardn/HERWIG/}
\url{ISAWIG/}

\bibitem{Corcella:2000bw}
  G.~Corcella {\it et al.},
  JHEP {\bf 0101}, 010 (2001)
  [arXiv:hep-ph/0011363].

\bibitem{Corcella:2002jc}
  G.~Corcella {\it et al.},
  arXiv:hep-ph/0210213.
  
\bibitem{Moretti:2002eu}
  S.~Moretti, K.~Odagiri, P.~Richardson, M.~H.~Seymour and B.~R.~Webber,
  JHEP {\bf 0204}, 028 (2002)
  [arXiv:hep-ph/0204123].

\bibitem{Ovyn:2009tx}
  S.~Ovyn, X.~Rouby, V.~Lemaitre,
  [arXiv:0903.2225 [hep-ph]].

\bibitem{Cacciari:2008gp}
  M.~Cacciari, G.~P.~Salam, G.~Soyez,
  JHEP {\bf 0804 } (2008)  063.
  [arXiv:0802.1189 [hep-ph]].

\bibitem{Cacciari:2005hq}
  M.~Cacciari, G.~P.~Salam,
  Phys.\ Lett.\  {\bf B641 } (2006)  57-61.
  [hep-ph/0512210].

\bibitem{Aad:2011uv}
  G.~Aad {\it et al.}  [\texttt{ATLAS} Collaboration],
  Phys.\ Rev.\ D {\bf 83} (2011) 112006
  [arXiv:1104.4481 [hep-ex]];
  \idem
  arXiv:1112.4828 [hep-ex].
  
\bibitem{Chatrchyan:2011ar}
  S.~Chatrchyan {\it et al.}  [\texttt{CMS} Collaboration],
  Phys.\ Lett.\ B {\bf 703} (2011) 246
  [arXiv:1105.5237 [hep-ex]];
  \idem
 CMS PAS EXO-11-028,
  \url{https://cdsweb.cern.ch/record/1405702}.
  
\end{thebibliography}
\end{document}